\documentclass[12pt]{article}

\usepackage{epsfig,amssymb}
\catcode`\@=11
\def \be  {\begin{equation}}
\def \ee  {\end{equation}}
\def \beq  {\begin{equation}}
\def \eeq  {\end{equation}}
\def \ba  {\begin{eqnarray}}
\def \ea  {\end{eqnarray}}
\def \baa {\begin{eqnarray*}}
\def \eaa {\end{eqnarray*}}
\def \lab #1 {\label{#1}}

\newcommand\re[1]{(\ref{#1})}
\def \qqquad {\qquad\quad}

\def \matrix #1 {\left(\begin{array}{cc} #1 \end{array}\right)}

\def \tr {\mathop{\rm tr}\nolimits}

\def \e  {\mathop{\rm e}\nolimits}

\newcommand{\insertfig}[2]{\mbox{\epsfxsize=#1cm \epsfbox{#2.eps}}}
\newcommand\lr[1]{{\left({#1}\right)}}
\newcommand \widebar [1] {\overline{#1}}

\newcommand \vev [1] {\langle{#1}\rangle}

\newcommand \ket [1] {|{#1}\rangle}
\newcommand \bra [1] {\langle {#1}|}

\newcommand{\as}{\ifmmode\alpha_{\rm s}\else{$\alpha_{\rm s}$}\fi}
\newcommand{\bit}[1]{\mbox{\boldmath$#1$}}

\font\cmss=cmss12 
\def\inbar{\,\vrule height1.5ex width.4pt depth0pt}
\def\IC{\relax\hbox{$\inbar\kern-.3em{\rm C}$}}
\def\IZ{\relax{\hbox{\cmss Z\kern-.4em Z}}}
\def\IR{{\hbox{{\rm I}\kern-.2em\hbox{\rm R}}}}

\def\IP{{\hbox{{\rm I}\kern-.2em\hbox{\rm P}}}}
\def\II{\hbox{{1}\kern-.25em\hbox{l}}}

\def\numberbysection{\@addtoreset{equation}{section}
                     \def\theequation{\thesection.\arabic{equation}}}
\numberbysection

\begin{document}

\begin{titlepage}

\hfill\parbox{40mm}{\begin{flushleft} DOE/ER/40762-278 \\ [-2mm] UMD-PP\#03-042
\\ [-2mm] ITEP-TH-16/03 \\ [-2mm] IHES/P/03/18 \\ [-2mm] LPT-Orsay-03-18
\end{flushleft}}

\vspace{1mm}

\centerline{\large \bf Gauge/string duality for QCD conformal operators}

\vspace{7mm}

\centerline{\bf A.V. Belitsky$^a$, A.S. Gorsky$^{b,c}$, G.P. Korchemsky$^d$}

\vspace{3mm}

\centerline{\it $^a$Department of Physics, University of Maryland at College Park}
\centerline{\it College Park, MD 20742-4111, USA}

\vspace{3mm}

\centerline{\it $^b$Institute of Theoretical and Experimental Physics}
\centerline{\it B. Cheremushkinskaya 25, Moscow, 117259, Russia}

\vspace{3mm}

\centerline{\it $^c$Institut des Hautes Etudes Scientifique}
\centerline{\it Le Bois-Marie, Bures-sur-Yvette, F-91440, France}

\vspace{3mm}

\centerline{\it $^d$Laboratoire de Physique Th\'eorique\footnote{Unit\'e
                    Mixte de Recherche du CNRS (UMR 8627).},
                    Universit\'e de Paris XI}
\centerline{\it 91405 Orsay C\'edex, France}

\vspace{5mm}

\centerline{\bf Abstract}

\vspace{5mm}

Renormalization group evolution of QCD composite light-cone operators, built from
two and more quark and gluon fields, is responsible for the logarithmic scaling
violations in diverse physical observables. We analyze spectra of anomalous
dimensions of these operators at large conformal spins at weak and strong
coupling with the emphasis on the emergence of a dual string picture. The
multi-particle spectrum at weak coupling has a hidden symmetry due to
integrability of the underlying dilatation operator which drives the evolution.
In perturbative regime, we demonstrate the equivalence of the one-loop cusp
anomaly to the disk partition function in two-dimensional Yang-Mills theory which
admits a string representation. The strong coupling regime for anomalous
dimensions is discussed within the two pictures addressed recently, --- minimal
surfaces of open strings and rotating long closed strings in AdS background. In
the latter case we find that the integrability implies the presence of extra
degrees of freedom -- the string junctions. We demonstrate how the analysis of
their equations of motion naturally agrees with the spectrum found at weak
coupling.

\vspace{2mm}

\noindent PACS numbers: 11.25.Hf, 11.15.Pg, 12.38.Cy

\end{titlepage}

\newpage

\tableofcontents

\newpage

\section{Introduction}

The formalism of path-ordered exponentials, or Wilson loops, is an indispensable
tool in QCD. It allows one to formulate complicated QCD dynamics in terms of
gauge invariant degrees of freedom \cite{Pol80} and express correlation functions
as a sum over random walks, e.g.,
\be
\vev{0|J_\mu(x) J_\nu(0)|0}
=
\sum_C \e^{- m \, L[C]} {\mit\Phi}_{\mu\nu}[C]
\ \vev{0| \tr P \exp \lr{i\oint_C dx_\mu A^\mu(x)} |0}
\, ,
\label{corr-func}
\ee
where $J_\mu(x) = \widebar {\mit\Psi} (x) \gamma_\mu {\mit\Psi} (x)$ is the
electromagnetic current of a quark with mass $m$. $L[C]$ is the length of a
closed path $C = C [0,x]$ that passes through the points $x$ and $0$.
${\mit\Phi}_{\mu\nu}[C]$ is a geometrical phase, the so-called Polyakov spin
factor, that takes into account the variation of the quark spin upon parallel
transport along the path $C$. To evaluate \re{corr-func}, one has to calculate
the (nonperturbative) expectation value of the Wilson loop for an arbitrary path
$C$ and perform resummation in the right-hand side of \re{corr-func}. Both tasks
are extremely difficult and can not be performed in full at the current stage.
Recently, a significant progress has been achieved in understanding the strong
coupling dynamics of supersymmetric gauge theories \cite{Mal98,EriSemZar00}
based on the gauge/string correspondence \cite{Mal97,GunKlePol98,Wit98}. One of
the goals of the present paper is to establish a relation between certain QCD
observables and their counterparts in string theory.

There exists a special class of QCD observables, for which the sum
over paths in the right-hand side of \re{corr-func} can be
performed exactly. As a relevant physical example, let us consider
a propagation of an energetic quark through a cloud of soft
gluons. In the limit when its energy goes to infinity, the quark
behaves as a point-like charged particle that moves along a straight
line and interacts with soft gluons. This means that the sum over
all paths in \re{corr-func} is dominated in that case by a saddle
point describing a propagation of a quark along its classical
path. The Wilson loop corresponding to this path has the meaning
of the eikonal phase acquired by the quark field upon interaction
with gluons. In this way, the Wilson loop encodes universal
features of soft radiation in QCD. Let us point out two important
QCD observables, in which similar semiclassical regime occurs: the
Isgur-Wise heavy-meson form factor, $\xi(\theta)$, and parton
distributions in a hadron, $f(x)$, at the edge of the phase space,
$x\to 1$. As we will demonstrate below, both observables are given
by an expectation value of a Wilson loop with the integration
contour $C$ fixed by the kinematics of the process. A unique
feature of the contour $C$ is that it contains a few cusps at points
in Minkowski space-time where the interaction with a large momentum
has occured in the underlying process.

In this way, Wilson loops with cusps, being fundamental objects in gauge
theories, have a direct relevance for QCD phenomenology. Their calculation in the
strong coupling (nonperturbative) regime is one of the prominent problems in
gauge theories. In the present paper, we make use of a recent progress in
understanding the strong coupling behavior of the ${\cal N} = 4$ supersymmetric
(SUSY) Yang-Mills (YM) theory to get some insights into properties of Wilson
loops in QCD. Our analysis relies on the gauge/string duality between
${\cal N} = 4$ SUSY gauge theory and a string theory on AdS$_5\times$S$^5$
background \cite{Mal97,GunKlePol98,Wit98}.%
\footnote{Note that recently there were several studies which aimed on the
derivation of strong coupling results for high-energy QCD observables, most
notably Refs.~\cite{GubKlePol02,DruGroOog99,Kru02,Mak02,JanPes00,PolSus01,GorKogKor02}.}
A natural question arises: what is in common between QCD and ${\cal N} = 4$
SUSY Yang-Mills theory? The two theories have quite different dynamics at
large distances, while at short distances they have many features in common.
For instance, anomalous dimensions of twist-two operators, contributing to
high-energy QCD processes, have a similar form in two theories including their
behavior at large Lorentz spin. Having this relation in mind, we will study
expressions for resummed anomalous dimensions in the ${\cal N} = 4$
SUSY Yang-Mills theory.%
\footnote{Since the anomalous dimensions originate from short distances,
we find it appropriate to refer to them in the strong coupling regime as
resummed anomalous dimensions rather then nonperturbative ones.}

We concentrate on two observations relevant to our present discussion. Recently,
it was proposed that anomalous dimensions of twist-two composite operators with
large Lorentz spin $J$ are equal in the strong coupling limit to the ``Energy
$-$ Spin'' of  a folded closed string rotating in AdS${}_5$ and having the shape
of a long rigid rod (mimicking the adjoint QCD string of glue with heavy quarks
at the folding points) \cite{GubKlePol02}
\be
\gamma_J(\alpha_s) = E - J
= 2 \sqrt{\frac{\alpha_s N_c}{\pi}} \, \ln J +
\mathcal{O}(J^0)
\, .
\label{GKP}
\ee
Another observation comes from the calculation of the Wilson loop in the strong
coupling regime via the minimal surface, $\mathcal{A}_{\rm min}$, swept by an
open string which goes into the fifth AdS dimension and whose ends trace its
contour in Minkowski space \cite{Mal98}. This picture naturally embeds the color
flux tubes between the color sources, albeit penetrating into an extra Liouville
dimension \cite{Pol98} as compared to the conventional four-dimensional setup.
{}From the QCD perspective, one is mostly interested in calculating Wilson loops
with cusps. Such contours were considered in a number of studies
\cite{DruGroOog99,Kru02,Mak02}. For the ${\scriptstyle \Pi}$-shaped Wilson loop
with two cusps (see Eq.~\re{W-DIS} below) the result reads \cite{Kru02,Mak02}
\be
W( v\!\cdot\!n \,\xi \mu)
=
\exp(i {\cal A}_{\rm min})
\, , \qquad
{\cal A}_{\rm min}
=
i \sqrt{\frac{\alpha_s N_c}{\pi}} \ln^2 (i v\!\cdot\!n \, \xi\mu)
\, .
\label{W-min}
\ee
Comparing \re{GKP} and \re{W-min}, one notices that $\gamma_J$ and $A_{\rm min}$
depend on the coupling constant in the same manner. The coincidence is not
accidental, of course. Identifying $\sqrt{\alpha_s}$-factor as the leading term
in the expression for the {\sl cusp anomalous dimension}
\be
{\mit\Gamma}_{\rm cusp} (\alpha_s) = \sqrt{\frac{\alpha_s N_c}{\pi}} +
\mathcal{O}((\sqrt{\alpha_s})^0)
\, ,
\label{cusp-strong}
\ee
one can show that Eqs.~\re{GKP} and \re{W-min} hold in a conformal gauge
theory for arbitrary coupling constant~\cite{KorMar93}. Eq.~\re{cusp-strong}
defines the asymptotic behavior of ${\mit\Gamma}_{\rm cusp} (\alpha_s)$ in
the $\mathcal{N} = 4$ SUSY YM theory in the strong coupling regime.

In the present paper, we will extend these results to composite QCD operators
of higher twist, built from an arbitrary number of fields and having autonomous
renormalization scale evolution. Such operators are known in QCD as
multi-particle conformal operators. We determine the spectrum of their anomalous
dimensions at large Lorentz spins, $J \gg 1$, both in weak and strong coupling
regimes. At weak coupling, the spectrum has a hidden symmetry due to
integrability of the dilatation operator in the underlying Yang-Mills theory. At
strong coupling, the two different pictures, --- the rotating folded long string
and the minimal surface swept by an open string with ends attached to the cusp,
--- result into the same asymptotic expression for the anomalous dimension of
conformal operators. We argue that integrability at weak coupling implies the
presence of extra stringy degrees of freedom at strong coupling, --- the string
junctions, --- and elucidate the relation between the anomalous dimensions of
multi-particle conformal operators at strong coupling and solutions to the
classical equations of motion for the string junctions.

To sew together the expressions for the anomalous dimensions at weak and strong
coupling, one needs the stringy description of the weak coupling regime in
Yang-Mills theory. One approach to the derivation of such stringy picture, based
on the hidden integrability of evolution equations for the light-cone operators,
has been developed in \cite{GorKogKor02}. It relies on the identification of the
underlying Yang-Mills dilatation operator as the Hamiltonian of
$SL(2,\mathbb{R})$ Heisenberg spin chain. The $SL(2,\mathbb{R})$ group naturally
appears in this context as a subgroup of the four-dimensional conformal group
acting on the light-cone. Due to complete integrability of the spin chain model,
the spectrum of the anomalous dimensions of multi-particle light-cone operators
can be found exactly in terms of the Riemann surfaces whose genus is related to
the number of the particles involved. As a consequence, the twist expansion on
the light-cone was shown to correspond to the summation over the genera of the
corresponding Riemann surfaces.

Our consequent presentation is organized as follows. In section~2 we review the
relation of certain QCD observables to expectation values of Wilson lines and
elucidate the physical meaning of the cusp anomaly. We also give there results
for the two-loop cusp anomalous dimensions in supersymmetric theories. In section
\ref{ConformalOp}, we turn to multi-particle operators and show how conformal
symmetry in gauge theory simplifies the problem of finding the spectrum of their
anomalous dimensions. We demonstrate the way the integrability of the evolution
equations arises through the identification of the underlying dilatation operator
with the Hamiltonian of a Heisenberg spin chain. In the subsequent section we
address the stringy interpretation of the gauge theory results. Our analysis
suggests that the cusp anomaly at weak coupling is described by a string which is
different from the Nambu-Goto string. To identify this string we show that the
cusp anomalous dimension to one-loop order is equal to the transition amplitude
for a test particle on the $SL(2,\mathbb{R})$ group manifold, which in its turn
is given by the partition function of two-dimensional Yang-Mills theory on a disk
which admits a stringy representation. Next, in section \ref{StrongCoupling}, we
discuss the strong coupling computation within the open and closed string theory
context for multi-particle operators extending earlier results. Section~6
contains concluding remarks. In the appendix, we calculate the contribution of
vacuum polarization to the two-loop cusp anomaly in dimensional regularization
and dimensional reduction schemes.

\section{Wilson loops as QCD observables}
\label{QCDptTheory}

As was emphasized in the introduction, there are several QCD
observables directly related to expectation values of Wilson
lines.

\subsection{Isgur-Wise form factor}

The Isgur-Wise form factor $\xi(\theta)$ describes the electromagnetic transition
of a heavy meson $|M (v) \rangle$ with mass $m$ and momentum $p_\mu \equiv m
v_\mu$, built from a heavy quark and a light component, to the same meson with
momentum $p'_\mu \equiv m v'_\mu$ (with $v_\mu^2 = {v'_\mu}^2 = 1$)
\cite{IsgWis90}
\begin{equation}
\label{IsgurWise}
\langle M (v') |
\widebar {\mit\Psi} (0) \gamma_\mu {\mit\Psi} (0)
| M (v) \rangle
=
\xi ( \theta ) (v + v')_\mu
\, .
\end{equation}
In the heavy-quark limit, $m\to \infty$, it depends only on the product of
velocities $v' \!\cdot\! v = \cosh \theta$, or equivalently on the angle
$\theta$ between them in Minkowski space-time. The operator ${\mit\Psi} (0)$
annihilates the heavy quark inside the meson $| M (v) \rangle$. For $m \to
\infty$, the heavy quark behaves as a classical particle with the velocity
$v_\mu$ interacting with the light component of the meson through its eikonal
current, $J^{a,{\rm eik}}_\mu (x) = \int_{-\infty}^0 d \tau\, v_\mu t^a
\delta^{(4)}(x-v\tau)$ with $t^a$ being the quark color charge. This allows one
to replace
\be
{\mit\Psi} (x) \rightarrow \e^{-im (v\cdot x)} b_v \, {\mit\Phi}_v [x; -
\infty]
\, , \qquad
{\mit\Phi}_v [x; - \infty] \equiv P \exp \left( i \int_{-\infty}^0 d \tau \, v
\cdot A (x + v \tau)
\right)\,,
\label{Phi-app}
\ee
where ${\mit\Phi}_v [x;-\infty]$ is the eikonal phase of a heavy quark in the
fundamental representation of the $SU(N_c)$ and $b_v$ amputates this quark inside
the heavy meson. Applying similar transformation to the quark field in the final
state meson $| M (v')
\rangle$, one obtains the following expression for the form factor
\cite{KorRad93}
\begin{equation}
\xi ( \theta )
=
\langle \widetilde{M}(v') |
{\mit\Phi}_{v'} [ \infty ; 0 ] {\mit\Phi}_v [ 0 ; - \infty ]
| \widetilde{M}(v) \rangle
\equiv
\vev{P \exp\lr{i \oint_{\wedge} dx_\mu A^\mu(x)}}
\,,
\label{W}
\end{equation}
with $| \widetilde{M}(v) \rangle = b_v | M (v) \rangle$ standing for the light
component of the meson with the amputated heavy quark. Here the net effect of
nonperturbative interaction with the light component of the heavy meson is
accumulated only via the Wilson line%
\footnote{Representing the ``brown-muck" of the heavy meson by a wave
function $| M_\ell \rangle = \int d^3 k \, \phi (k) \, a^\dagger_k | 0
\rangle$, the transition of the light cloud from the initial to the
final state can be described by an exact light-quark propagator in an
external gluon field. In this manner the Isgur-Wise form factor will be
rewritten as a correlation function of Wilson loop along the contour formed
by the straight-line trajectories of heavy quarks and the phase of the light
spectator quark in the world-line expression of its propagator \cite{Pol88}.}
evaluated along the contour consisting of two rays that run along the meson
velocities $v_\mu$ and $v'_\mu$. It is important to notice that the contour
has a cusp at the point $0$, in which the interaction with the external probe
has occurred.

The Isgur-Wise form factor $\xi(\theta)$ is a nonperturbative observable in QCD.
It depends on hadronic, long-distance scales as well as on the ultraviolet
cut-off $\mu\sim m$, which sets up the maximal energy of soft gluons. Although
$\xi(\theta)$ can not be calculated at present in QCD from the first principles,
its dependence on $\mu$ can be found from the renormalization group equation
\begin{equation}
\left(
\mu \frac{\partial}{\partial\mu}
+
\beta (g) \frac{\partial}{\partial g}
+ {\mit\Gamma}_{\rm cusp} (\theta; \alpha_s)
\right) \xi ( \theta ) = 0 \, ,
\label{RG}
\end{equation}
where $\alpha_s = g^2/(4 \pi)$ is the QCD coupling constant and
${\mit\Gamma}_{\rm cusp} (\theta; \alpha_s)$ is the cusp anomalous dimension
\cite{Pol80,KorRad86}. To the lowest order in $\alpha_s$
\begin{equation}
\label{GammaCusp}
{\mit\Gamma}_{\rm cusp} (\theta; \alpha_s) =
\frac{\alpha_s C_F}{\pi}
\left( \theta \coth \theta - 1 \right)
+ \mathcal{O}(\alpha_s^2)
\, ,
\end{equation}
where $C_F = (N_c^2 - 1)/(2N_c)$ is the Casimir operator of the $SU(N_c)$ group
in the fundamental representation. The two-loop correction to \re{GammaCusp} has
been calculated in \cite{KorRad86} and its dependence on $\theta$ is more involved.

Eq.\ \re{RG} follows from renormalization properties of the Wilson
line in the right-hand side of \re{W}. It acquires the anomalous
dimension due to the presence of a cusp on the integration
contour. The cusp anomalous dimension ${\mit\Gamma}_{\rm cusp}
(\theta; \alpha_s)$  determines universal features of soft-gluon
radiation and is known as the {\sl QCD bremsstrahlung function\/}.
As such, it is a positive definite function of the cusp angle (for
real Minkowski angle $\theta$) at arbitrary value of the coupling
constant
\be
{\mit\Gamma}_{\rm cusp} (\theta; \alpha_s)
\ge 0\,.
\label{pos}
\ee
To see this we recall that at the cusp point the heavy quark
suddenly changes its velocity from $v_\mu$ to $v'_\mu$ and, due to
instantaneous acceleration, it starts to emit  soft (virtual and
real) gluons with momentum $k<\mu$ with a cut-off $\mu\sim m$.
Denoting the eikonal phase of the heavy quark as ${\mit\Phi}
\equiv {\mit\Phi}_{v'} [\infty ; 0 ] {\mit\Phi}_v [ 0 ; - \infty
]$ and using its unitarity, one calculates the total probability
for the heavy quark to undergo the scattering (the Bjorken sum
rule) as
\be
1
=
\vev{ \widetilde{M}(v)| {\mit\Phi}^\dagger {\mit\Phi} |\widetilde{M}(v)}
=
|\xi (\theta)|^2
+
\sum_X
\left|
\vev{\widetilde{M}_X(v')| {\mit\Phi} |\widetilde{M}(v)}
\right|^2
\, ,
\label{unitarity}
\ee
where in the right-hand side we inserted the decomposition of the unity operator
over the physical hadronic states and separated the contribution of the ground
state meson, $\ket{\widetilde{M}(v)}$, from excited states
$\ket{\widetilde{M}_X(v)}$. The Wilson line \re{W} defines the probability of the
elastic transition, $|\xi|^2\sim\exp(-w)$ with $w = 2\int^\mu (d k/k) \,
{\mit\Gamma}_{\rm cusp} (\theta; \alpha_s(k))$. For $\theta\neq 0$, depending on
the sign of ${\mit\Gamma}_{\rm cusp}(\theta;\alpha_s)$, it either vanishes or
goes to infinity for $\mu \to \infty$. In order to preserve the unitarity
condition $|\xi|^2 \le 1$ that follows from \re{unitarity}, one has to require
that $\xi\to 0$ for $\mu \to \infty$ leading to \re{pos}. At $\theta = 0$ the
cusp vanishes, that is the heavy meson stays intact, the sum in \re{unitarity}
equals zero and $\xi(\theta=0) = 1$. This implies that the cusp anomalous
dimension vanishes for $\theta \to 0$.

\subsection{Parton distributions at $x \to 1$}

Our second example is provided by deeply inelastic scattering of a hadron $H(p)$
with momentum $p_\mu$ off a virtual photon $\gamma^*(q)$ with momentum $- q_\mu^2
= Q^2\gg p^2$ in the exclusive limit $x_{\rm Bj} = Q^2/(2p \cdot q) \to 1$, i.e.\
when the invariant mass of the final state system becomes small $(q + p)^2
\ll Q^2$. In the scaling limit, $Q^2 \to \infty$, the cross
section of the process is expressed in terms of the twist-two
quark distribution function \cite{ColSop82}, see also Ref.\
\cite{BelJiYua02},
\begin{equation}
\label{CollinearPDFs}
f (x)
=
\int_{-\infty}^\infty \frac{d \xi}{2 \pi} \,
{\rm e}^{- i \, x \, \xi}
\langle H(p) |
\widebar {\mit\Psi} (\xi n) {\mit\Gamma} {\mit\Phi}_n [ \xi ; 0 ] {\mit\Psi} (0)
| H(p) \rangle
\, ,
\end{equation}
describing the probability to find a quark inside the hadron with the fraction
$x$ of its momentum $p$. The Wilson line stretched in between the quark fields
makes the bilocal operator gauge invariant. It goes along the light-like
direction $n_\mu = (q_\mu +p_\mu x_{\rm Bj})/(p\!\cdot\!q)$, so that $n^2=0$ and
$n\!\cdot\!p = 1$.

The matrix ${\mit\Gamma} = {\not\!n}$ in \re{CollinearPDFs} serves to select the
quark states with opposite helicities. For our purposes, we will not specify
${\mit\Gamma}$ and treat it as a free parameter. The Mellin moments of the
distribution function \re{CollinearPDFs} are related to the matrix elements of
local twist-two operators
\begin{equation}
\int_0^1 d x \, x^J f (x; \mu^2)
= \langle H(p) |
\widebar {\mit\Psi} (0)
{\mit\Gamma} \left( i \, n \!\cdot\! {\cal D} \right)^J
{\mit\Psi} (0)
| H(p) \rangle
\equiv \langle {\cal O}_J^{\mit\Gamma} (\mu^2)\rangle
\, ,
\label{O}
\end{equation}
where ${\cal D}_\mu=\partial_\mu-iA_\mu$ is a covariant derivative. Their
dependence on the ultraviolet cut-off $\mu$ is described by an evolution
equation, whose solution reads in terms of the anomalous dimensions
\begin{equation}
\langle {\cal O}_J^{\, \mit\Gamma} (\mu^2)  \rangle
=
\left( \mu / \mu_0 \right)^{-\gamma_J^{\,\mit\Gamma} (\alpha_s)}
\langle {\cal O}_J^{\,\mit\Gamma} (\mu_0^2) \rangle
\,,
\label{O-sol}
\end{equation}
where we assumed for simplicity that the coupling constant does not run, $\beta =
0$. The anomalous dimension of the twist-two operators, $\gamma_J^{\mit\Gamma}
(\alpha_s)$, depends on the choice of the matrix ${\mit\Gamma}$. In particular,
in case when ${\mit\Gamma}$ selects the same helicities of the quark fields in \re{O},
${\mit\Gamma} = (1 + \gamma_5) {\not\!n} \gamma_\perp$, the anomalous dimension is
\begin{equation}
\label{AnomalousDimension}
\gamma_J(\alpha_s)
=
\frac{\alpha_s}{\pi} C_F
\Big( 2 \psi (J + 2) + 2 \gamma_{\rm E} - 3/2 \Big)
+ \mathcal{O}(\alpha_s^2)
\, ,
\label{psi-func}
\end{equation}
where $\psi (J) = d \ln {\mit\Gamma} (J)/d J$ is the Euler
psi-function and $\gamma_{\rm E}$ is the Euler constant. For other
choices of ${\mit\Gamma}$, the anomalous dimensions have extra
(rational in $J$) terms in addition to the $\psi$-function, see,
e.g., \cite{BukFroKurLip85}. As we will argue below, Eq.\
\re{psi-func} has a hidden symmetry which is responsible for
integrability of evolution equations for three-quark (baryonic)
composite operators. Going over to the $\mathcal{N} = 4$ SUSY
Yang-Mills theory, one finds that the same expression
\re{psi-func} defines (up to redefinition of the color factor $C_F
\to N_c$) the anomalous dimensions of multiplicatively
renormalizable twist-two operators \cite{KotLip}.

As follows from \re{O}, the asymptotics of the distribution function for
$x \to 1$ is related to the contribution of twist-two operators of large
Lorentz spins $J \sim 1/(1 - x) \gg 1$. One finds from \re{AnomalousDimension}
that the anomalous dimension scales in this limit as
\be
\gamma_{J} (\alpha_s) = \frac{\alpha_s}{\pi} C_F
\left\{
\ln (J+2) + \gamma_{\rm E} - 3/4 - \frac{1}{2 (J+2)}
- \sum_{n = 1}^{\infty} \frac{B_{2n}}{2n} (J+2)^{- 2 n}
\right\}
+ \ldots
\,,
\label{gamma-exp}
\ee
where $B_{n}$'s are the Bernoulli numbers, $B_2 = 1/6$, $B_4 = -1/30, \, \dots$,
and the ellipsis stands for higher order terms in $\alpha_s$. It turns out that
the leading scaling behavior $\gamma_J^{\mit\Gamma} (\alpha_s) \sim \ln J$ is a
universal property of the anomalous dimensions of the twist-two operators \re{O}
for arbitrary ${\mit\Gamma}$. It holds to all orders in $\alpha_s$ and is
intrinsically related to the cusp anomaly of the Wilson loops. The reason for
this is that analyzing deeply inelastic scattering for $x \to 1$ one encounters
the same physical phenomenon as in the case of the Isgur-Wise form factor, i.e.\
the struck quark carries almost the whole momentum of the hadron and, therefore,
it interacts with other partons by exchanging soft gluons. In these circumstances,
in complete analogy to the previous case, Eq.\ \re{Phi-app}, the quark field
can be approximated by an eikonal phase evaluated along the classical path in
the direction of its velocity $p_\mu = m v_\mu$,
\begin{equation}
\label{LargeX}
f (x)
=
\int_{-\infty}^\infty \frac{d \xi}{2 \pi}
{\rm e}^{i(1 - x) \, \xi}
\langle \widetilde{H}(p) |
W_{\scriptstyle\Pi} ( {v \!\cdot\! n} \, \xi \mu -i0) | \widetilde{H}(p) \rangle
\, ,
\end{equation}
where $| \widetilde{H}(p) \rangle$ is the state of the target hadron with
amputated energetic quark and the causal $-i0$ prescription ensures the correct
spectral property, $f (x) = 0$ for $ x > 1$. The ${\scriptstyle \Pi}$-shaped
Wilson line in Eq.\ (\ref{LargeX}) consists out of two rays and one segment:
a link from $- \infty$ to $0$ along the velocity of the incoming quark, next
along the light-cone direction $n_\mu$ to the point $\xi n_\mu $ and, then,
along $- v_\mu$ from $0$ to $\infty$,
\begin{equation}
W_{\scriptstyle\Pi} ({v \!\cdot\! n} \, \xi\mu )
=
{\mit\Phi}^\dagger_v [ \xi ; - \infty ]\,
{\mit\Phi}_n [ \xi ; 0 ]\, {\mit\Phi}_v [ 0 ; - \infty ]
\, .
\label{W-DIS}
\end{equation}
Substituting \re{LargeX} into \re{O}, one finds the following relation between
the matrix elements of local composite operators at large spin $J$ and the Wilson
loop expectation value~\cite{KorMar93}
\begin{equation}
\langle{\cal O}_J^{\,\mit\Gamma} (\mu^2) \rangle
=
\langle \widetilde{H}(p) |
W_{\scriptstyle\Pi} ( - i J )
| \widetilde{H}(p)  \rangle
\equiv\vev{P\exp\lr{i\int_{\scriptstyle\Pi} dx_\mu A^\mu(x)}}
\, .
\label{O-W}
\end{equation}
Here, the large Lorentz spin of the local operator defines the length
of the light-cone segment:
\begin{equation}
\label{JtoTheta}
{v \!\cdot\! n} \, \xi \, \mu \to - i J
\, .
\end{equation}
We would like to stress that Eq.~\re{O-W} holds only for $J \gg 1$.

According to Eq.~\re{O-W}, the $\mu$-dependence of the twist-two operators
follows from the renormalization of the Wilson line \re{W-DIS}. The latter has
two cusps located at the points $0$ and $\xi n_\mu$. In distinction with the
previous case, one of the segments attached to the cusps lies on the light-cone,
$n^2 = 0$, and the corresponding cusp angle is infinite, $\theta\sim \frac12
\ln[ (v \!\cdot\! n)^2/n^2] \to \infty$. In this limit, the cusp anomalous
dimension scales to all orders in $\alpha_s$ as \cite{KorRad86}
\begin{equation}
{\mit\Gamma}_{\rm cusp} (\theta; \alpha_s) =
\theta \, {\mit\Gamma}_{\rm cusp} (\alpha_s) + \mathcal{O}(\theta^0)
\,.
\label{light-cusp}
\end{equation}
Here ${\mit\Gamma}_{\rm cusp}(\alpha_s)$ is a universal anomalous dimension
independent on $\theta$. At weak coupling, it has the following form in QCD
\be
{\mit\Gamma}_{\rm cusp}(\alpha_s)
=
\frac{\alpha_s}{\pi}C_F
+
\lr{\frac{\alpha_s}{\pi}}^2 C_F
\left\{
N_c \lr{\frac{67}{36} - \frac{\pi^2}{12}} - n_f \frac5{18}\right\}
+
\mathcal{O}(\alpha_s^3)
\,,
\label{two-loop}
\ee
where $n_f$ is the number of quark flavors. This expression was obtained
within the dimensional regularization scheme (DREG) by using the
$\widebar{\textrm{MS}}$-subtraction procedure, $\alpha_s \equiv
\alpha_s^{_{\widebar{\rm MS}}}$.

The divergence of the anomalous dimension \re{light-cusp} for $\theta\to\infty$
indicates that the Wilson line with a light-like segment satisfies an evolution
equation different from \re{RG}. The modified equation looks like \cite{KorMar93}
\begin{equation}
\left(
\mu \frac{\partial}{\partial\mu}
+
\beta (g) \frac{\partial}{\partial g}
+
2 {\mit\Gamma}_{\rm cusp} (\alpha_s) \ln[ i (v \!\cdot\! n) \, \xi \mu]
+
{\mit\Gamma} (\alpha_s)
\right)
\vev{ W_{\scriptstyle\Pi} ( {v \!\cdot\! n} \, \xi \mu) }
=
0
\, .
\label{RG-light}
\end{equation}
Here the factor of 2 stems from the presence of two cusps on the ${\scriptstyle
\Pi}$-shaped line contour and ${\mit\Gamma} (\alpha_s)$ is a process-dependent
anomalous dimension. The explicit dependence of the anomalous dimensions on
the renormalization scale $\mu$ implies the absence of the multiplicative
renormalizability of the light-like Wilson line. Combining together Eqs.\
\re{RG-light} and \re{O-W}, we obtain the renormalization group equation for
local composite operators $\langle{\cal O}_J^{\,\mit\Gamma}(\mu^2) \rangle$
at large $J$. Matching its solution into \re{O-sol}, we find the asymptotic
behavior of the anomalous dimensions of the twist-two quark operators for
$J \to \infty$
\begin{equation}
\gamma_J^{(qq)} (\alpha_s)
= 2 {\mit\Gamma}_{\rm cusp} (\alpha_s) \ln J + {\cal O} (J^0)
\,.
\label{gamma-as}
\end{equation}
Repeating a similar analysis for the twist-two gluon operators, one can show that
their matrix elements satisfy \re{O-W} with the Wilson line defined in the
adjoint representation. Therefore, their anomalous dimension satisfies
\re{gamma-as} upon replacing $C_F\to N_c$ leading to~\cite{KorMar93}
\be
\gamma_J^{(gg)} (\alpha_s)
=
\frac{N_c}{C_F} \, \gamma_J^{(qq)} (\alpha_s)
+
{\cal O} (J^0)
\, .
\ee
In general, the quark and gluon operators mix with each other. However, at large
$J$ the mixing occurs through the exchange of a soft quark with momentum $\sim
1/J$. Its contribution to the corresponding anomalous dimensions is suppressed by
a power of $1/J$ leading to
\be
\gamma_J^{(gq)} (\alpha_s) = {\cal O}(1/J)
\, , \qquad
\gamma_J^{(qg)} (\alpha_s) = {\cal O}(1/J)
\, .
\label{off-diag}
\ee
We would like to stress that the relations \re{gamma-as}--\re{off-diag} are valid
to all orders in $\alpha_s$. Remarkably enough, they hold both in QCD and its
supersymmetric extensions. In the latter case, the mixing matrix has a bigger
size due to the presence of additional scalar fields. Nevertheless, this matrix
remains diagonal at large $J$. Since the fields in supersymmetric YM theories
belong to the adjoint representation, the diagonal matrix elements are the same
\be
\gamma_J^{(ab)}
=
2 \delta_{ab} \mit{\Gamma}_{\rm cusp}(\alpha_s) \ln J
+
\mathcal{O}(J^0)
\, .
\ee
with $a,b=(q,g,s)$. However, one can not use for
$\mit{\Gamma}_{\rm cusp}(\alpha_s)$ the two-loop expression
\re{two-loop} at $C_F=N_c$, because it was obtained within the
dimensional regularization scheme which breaks supersymmetry and
also lacks the contribution of possible scalars.

\subsection{Cusp anomaly in supersymmetric theories}

To calculate ${\mit\Gamma}_{\rm cusp}(\alpha_s)$ in a supersymmetric Yang-Mills
theory, we use the regularization by dimensional reduction (DRED)
\cite{Sie79,CapJonNie80,Sie80}. In analogy to compactification, one gets this
scheme by dimensionally reducing the four-dimensional theory down to $d = 4 - 2
\varepsilon < 4$ dimensions. In comparison with the DREG scheme, the Lagrangian
involves now the so-called epsilon-scalars generated by $2 \varepsilon$
components of the four-dimensional gauge field. To two-loop accuracy, these
scalar fields contribute to the Wilson loop by modifying the self-energy of a
gluon at the level of $\mathcal{O}(\varepsilon)$ corrections. The calculation of
the corresponding Feynman diagram is straightforward and details can be found in
the Appendix \ref{DRED}. It leads to the two-loop correction $-
\lr{{\alpha_s^{{\rm \widebar{\scriptscriptstyle DR\!}}}}/{\pi}}^2 C_F N_c/12$ to
the right-hand side of \re{two-loop}, resulting into
\be
{\mit\Gamma}_{\rm cusp}
=
\frac{\alpha_s^{{\rm \widebar{\scriptscriptstyle DR\!}}}}{\pi} N_c
+
\left(
\frac{\alpha_s^{{\rm \widebar{\scriptscriptstyle DR\!}}}}{\pi}
\right)^2
N_c
\left\{
N_c \left( \frac{16}{9} - \frac{\pi^2}{12} \right)
-
n_f \frac{5}{18} - n_s \frac{1}{9}
\right\}
+
\mathcal{O}
\left(
(\alpha_s^{{\rm \widebar{\scriptscriptstyle DR\!}}})^3
\right)
\, ,
\label{two-loop-DR}
\ee
where $\alpha_s=\alpha_s^{{\rm \widebar{\scriptscriptstyle DR\!}}}$ is the
coupling constant in the DRED scheme with modified minimal subtractions. Here, in
comparison with \re{two-loop}, we added the contribution of $n_s = N_s N_c$ real
scalars and set $C_F = N_c$ since in a supersymmetic gauge theory fields belong
to the adjoint representation of the $SU (N_c)$. Notice that one can obtain the
same expression \re{two-loop-DR} by expanding \re{two-loop} in powers of the
coupling constant in the dimensional reduction scheme, $\alpha_s^{{\rm \widebar{
\scriptscriptstyle DR\!}}}$, which is related to the coupling constant in the
dimensional regularization scheme, $\alpha_s^{{\rm \widebar{\scriptscriptstyle
MS\!}}}$, by a
finite renormalization \cite{AltCurMarPet81,SchSakKor87}%
\footnote{This simple ``rule of substitutions" can be understood
by noticing that ${\mit\Gamma}_{\rm cusp}$ governs the scale
dependence of a physical observable, the Isgur-Wise form factor,
and, therefore, it is renormalization scheme invariant.}
\be
\label{MStoDR}
\alpha_s^{{\rm \widebar{\scriptscriptstyle MS\!}}}
=
\alpha_s^{{\rm \widebar{\scriptscriptstyle DR\!}}}
\lr{
1
-
\frac{N_c}{12} \frac{\alpha_s^{{\rm \widebar{\scriptscriptstyle DR\!}}}}{\pi}
+
\mathcal{O}
\left(
(\alpha_s^{{\rm \widebar{\scriptscriptstyle DR\!}}})^2
\right)}
\, .
\ee
Eqs.~\re{two-loop} and \re{two-loop-DR} define the cusp anomalous
dimensions in two different renormalization schemes, based on
dimensional regularization and dimensional reduction, respectively.
It is the latter scheme that does not break supersymmetry.

Using \re{gamma-as} and \re{two-loop-DR} we obtain the asymptotic behavior
of the twist-two anomalous dimension in the $\widebar{\rm DR}$ scheme
($\alpha_s \equiv \alpha_s^{{\rm \widebar{\scriptscriptstyle DR\!}}}$) in
various supersymmetric theories:

$\bullet$ In the $\mathcal{N}=1$ YM theory, one has one Majorana
fermion in the adjoint representation, $n_f = N_c$, and no scalars
$n_s=0$
\be
{\mit\Gamma}^{\mathcal{N} = 1}_{\rm cusp} (\alpha_s)
=
\frac{\alpha_s N_c}{\pi}
+
\lr{\frac{\alpha_s N_c}{\pi}}^2
\lr{\frac{3}{2} - \frac{\pi^2}{12}}
+
\mathcal{O}(\alpha_s^3)
\,.
\label{N1}
\ee

$\bullet$ In the $\mathcal{N} = 2$ YM theory, one has two Majorana
fermions in the adjoint representation, $n_f = 2 N_c$, and two
real scalar fields in the adjoint representation, $n_s = 2 N_c$
\be
{\mit\Gamma}^{\mathcal{N} = 2}_{\rm cusp} (\alpha_s)
=
\frac{\alpha_s N_c}{\pi}
+
\lr{\frac{\alpha_s N_c}{\pi}}^2 \lr{1-\frac{\,\pi^2}{12}}
+
\mathcal{O}(\alpha_s^3)
\, .
\label{N2}
\ee

$\bullet$ In the $\mathcal{N}=4$ YM theory, one has four Majorana
fermions in the adjoint representation, $n_f = 4N_c$, and  six
real scalar fields in the adjoint representation, $n_s = 6N_c$
\be
{\mit\Gamma}^{\mathcal{N} = 4}_{\rm cusp} (\alpha_s)
=
\frac{\alpha_s N_c}{\pi}
+
\lr{\frac{\alpha_s N_c}{\pi}}^2\lr{-\frac{\,\pi^2}{12}}
+
\mathcal{O}(\alpha_s^3)\,.
\label{N4}
\ee
Notice that the two-loop correction to ${\mit\Gamma}_{\rm cusp}(\alpha_s)$
is positive in all cases except the $\mathcal{N}=4$ theory. It becomes
smaller as one goes from QCD to the $\mathcal{N}=1$ and $\mathcal{N}=2$
theory and, then, it becomes negative at $\mathcal{N}=4$. We recall that
${\mit\Gamma}_{\rm cusp}(\alpha_s)> 0$ for arbitrary $\alpha_s$, Eq.~\re{pos}.

Together with Eqs.~\re{gamma-as}--\re{off-diag}, the expressions \re{N1}--\re{N4}
establish the large-$J$ asymptotics of the anomalous dimensions of the twist-two
operators in supersymmetric theories. Namely, the matrix of anomalous dimension
is diagonal at large $J$ with the entries on the main diagonal equal to $2
{\mit\Gamma}_{\rm cusp} (\alpha_s) \ln J$. Eq.~\re{N1} agrees with the results of
explicit two-loop calculations in Ref.~\cite{FloKeh}. Eq.~\re{N4} is in
disagreement with the
results of Ref.~\cite{KotLip}.%
\footnote{Eq.~\re{N4} were in agreement with the results of
Ref.~\cite{KotLip} if one would assume that their result is given in the
$\widebar{\rm MS}$ scheme and transforms the coupling constant to the
$\widebar{\rm DR}$ scheme via (\ref{MStoDR}).} Eq.~\re{N2} is a prediction since
a two-loop calculation in that case has not been performed yet.

\section{Conformal operators and integrable spin chains}
\label{ConformalOp}

So far we have discussed two-particle composite operators. Let us
now generalize our consideration to operators involving many
fields. Such operators are of great phenomenological interest as
their matrix elements define, e.g., baryon distribution amplitudes
\cite{LepBro80} and higher-twist corrections in various
high-energy processes \cite{Jaf97}. To start with, we define first
a framework which solves partially the expected complications in the
mixing problem in this case.

\subsection{Two-particle conformal operators}

The local twist-two operators $\mathcal{O}_{J}$ can be obtained from the
expansion of a nonlocal light-cone operator, cf.\ Eq.\ (\ref{CollinearPDFs}),
\be
\tr \left\{ X_1 (0) \, {\mit\Phi}_n [0, \xi] \, X_2 (\xi n) \,
{\mit\Phi}_{-n} [\xi, 0] \right\} = \sum_{J = 0}^\infty \frac{(- i
\xi)^J}{J!} \mathcal{O}_{J} (0) \, .
\label{gen-fun}
\ee
Here $X_{i}(n \xi)\equiv X_{i}^a(n \xi) t^a$ denotes a general primary operator
in the gauge theory defined in the adjoint representation of the $SU (N_c)$
group, ``living'' on the light-cone $n_\mu^2 = 0$ and having definite quantum
numbers with respect to transformations of the conformal group (see
Eq.~\re{Moebius} below). The latter condition implies that $X_{i}^a$ can be
identified as the quasi-partonic operator~\cite{BukFroKurLip85}, that is a scalar
field, or a specific component of the quark field and the gluon strength tensor.

Two Wilson lines in \re{gen-fun} run along the light-cone in opposite directions
between the points $0$ and $\xi$ and ensure the gauge invariance of the operator.
The local twist-two operators ${\cal O}_J$ were introduced in the previous
section. Let us reinstate their definition again,
\be
{\cal O}_J(\xi) = \tr\left\{ X_1 (n\xi) \left( i \, n \!\cdot\!
{\cal D} \right)^{J} X_2 (n\xi)\right\} = \tr\left\{ X_1 (\xi)
\left( i \partial \right)^{J} X_2 (\xi)\right\}
\, ,
\label{O-naive}
\ee
where in distinction with the previous case, Eq.~\re{O}, the covariant derivative
is defined in the adjoint representation, $\mathcal{D}_\mu
X=\partial_\mu-ig[A_\mu,X]$. Here in the second relation we have chosen the gauge
$n \cdot A (x) = 0$ and simplified the notation for the argument of the fields.
The twist-two operators defined in this way are not renormalized multiplicatively
and mix with operators containing total derivatives $(i\partial)^l{\cal
O}_{J-l}(\xi)$ with $1
\le l
\le J$. Although the mixing can be neglected for forward matrix elements like
\re{CollinearPDFs}, it survives in case one considers matrix elements with
different momenta in the initial and final state, or when they are a part of
multi-parton operators. In conformal theories, one can construct linear
combinations of such operators, the so-called conformal operators
\cite{BroLep79,EfrRad78,Mak80,Ohr82,BelMul98,Gor89}, in such a way that they
are renormalized autonomously to all orders in the coupling constant.%
\footnote{In gauge theories with broken conformal symmetry, like QCD, this
holds only to the lowest order in $\alpha_s$. Beyond the leading order
$\widehat O_J$'s start to mix with $(i \partial)^{l}\widehat O_{J - l}$. The
corresponding mixing anomalous dimensions are determined solely by the
conformal anomalies, see the last paper of Ref.\ \cite{BelMul98}, Eqs.\ (51)
and (113) for explicit expressions.} The mixing between these operators is
protected by the $SO(4,2)$ conformal symmetry of the gauge theory, more
precisely, by its collinear $SL(2,\mathbb{R}) \sim SU (1,1)$ subgroup, which
acts on the primary fields ``living'' on the light-cone as follows
\be
\xi \to \frac{a \xi + b}{c \xi + d}
\, , \qqquad
X_i(\xi) \to (c \xi + d)^{-2 j_i} X_i \lr{\frac{a \xi + b}{c \xi + d}}
\, ,
\label{Moebius}
\ee
with $a,\ldots,d$ real such that $ad - bc=1$. Here $j_i=(s_i+l_i)/2$ is the
conformal spin of the field $X_i(\xi)$ equal to one half of the sum of its
canonical dimension, $l$, and projection of the spin onto the light-cone,
$\Sigma_{-+} X_i(\xi) = s_i X_i(\xi)$. By definition, the conformal operators
$\widehat{\cal O}_J(\xi)$ are composite operators built from the primary fields
and satisfying \re{Moebius} with $j_i$ replaced by $J+j_1+j_2$. It is easy to see
that the operators \re{O-naive} do not obey the latter condition.

To simplify the analysis, one chooses the axial gauge $n \cdot A(x) = 0$.
Then, the operators in the left-hand side of \re{gen-fun} are given by the
product of two primary fields. It is transformed under \re{Moebius} in
accordance with the direct product of two $SL(2,\mathbb{R})$ representations%
\footnote{These are unitary representations of the $SL(2,\mathbb{R})$ group of
the discrete series.} labelled by the spins $j_1$ and $j_2$. Decomposing this
product into the sum of irreducible components, $[j_1]\otimes[j_2]=\sum_{J\ge 0}
[J+j_1+j_2]$, one can identify the spin-$J$ component as defining the conformal
operator $\widehat{\cal O}_J(\xi)$. It has the following form
\be
\widehat{\cal O}_J(\xi) = i^J \lr{\partial_2 + \partial_1}^J
P_J^{(\nu_1,\nu_2)} \lr{\frac{\partial_2 - \partial_1}{\partial_2
+ \partial_1}} \tr\left\{ X_1 (\xi_1)
X_2(\xi_2)\right\}\bigg|_{\xi_1 = \xi_2 = \xi} \,,
\label{Conf-op}
\ee
where $\partial_a = \partial/\partial \xi_a$, $\nu_a = 2 j_a - 1$ and
$P_J^{(\nu_1,\nu_2)}$ are the Jacobi polynomials. To restore gauge invariance,
one has to substitute $\partial X(\xi) = (n\cdot\mathcal{D})X(\xi)$. Going
back to \re{gen-fun}, one obtains the operator product expansion for a
nonlocal light-cone operator over the conformal operators, see, e.g., Refs.\
\cite{FerGatGri73,BalBra89,Mul98},
\be
\tr \left\{ X_1 (0) \, {\mit\Phi}_n [0, \xi] \, X_2 (\xi n) \,
{\mit\Phi}_{-n} [\xi, 0] \right\} = \sum_{J = 0}^\infty C_J
(\nu_1, \nu_2) \, \frac{(- i \xi)^J}{J!} \int_0^1 du\, u^{J +
\nu_1} (1 - u)^{J + \nu_2} \widehat{\cal O}_J (u\xi) \,,
\label{OPE-conf}
\ee
where
\begin{equation}
C_J (\nu_1, \nu_2)
=
(2 J  + \nu_1 + \nu_2 + 1)
\frac{
{\mit\Gamma} (J + 1) {\mit\Gamma} (J + \nu_1 + \nu_2 + 1)
}{
{\mit\Gamma} (J + \nu_1 + 1) {\mit\Gamma} (J + \nu_2 + 1)
}
\, .
\end{equation}
The Lorentz operators ${\cal O}_J$ can be re-expressed in terms of the conformal
operators $\widehat {\cal O}_J$ via
\begin{equation}
{\cal O}_J = \sum_{j = 0}^J c^J_j (\nu_1, \nu_2)
i^{J - j} \left( \partial_2 + \partial_1 \right)^{J - j} \widehat {\cal O}_j
\, ,
\end{equation}
with the expansion coefficients
$
c^J_j (\nu_1, \nu_2) = C_j (\nu_1, \nu_2)
\int_{0}^{1} d u \, (1 - u)^{\nu_1} u^{\nu_2 + J} P^{(\nu_1, \nu_2)}_j (2 u - 1)
\, ,
$
which can be calculated in terms of the hypergeometric function ${_3F_2}$. As
was already mentioned, the conformal operators evolve autonomously under
renormalization and their anomalous dimensions have a universal scaling
behavior at large $J$, Eqs.~\re{gamma-as} and \re{GKP}.

Let us consider the forward matrix element of the both sides of \re{OPE-conf} for
large light-cone separations, $\xi\gg 1$. In this limit, typical wavelength of
gluons exchanged between two fields $X_1(0)$ and $X_2(\xi n)$ scales as $\xi$ and
the eikonal approximation \re{Phi-app} is justified. This allows us to replace
the quantum operators $X_i(\xi)$ in the left-hand side of \re{OPE-conf} by Wilson
lines in the adjoint representations of the $SU(N_c)$. Then, the left-hand side
of \re{OPE-conf} can be approximated as $\vev{p\,|\widetilde W_\Pi(\xi)|p} \e^{-
i\xi (p \cdot n)}$, where the Wilson line in the adjoint representation,
$\widetilde W_{\scriptstyle\Pi} (\xi)$, evaluated along the same
${\scriptstyle\Pi}$-like contour as in \re{W-DIS}. In the right-hand side of
\re{OPE-conf}, one can neglect the contribution of operators with total
derivatives since their forward matrix elements vanish. In addition, at large
$\xi$ the sum is dominated by the contribution of operators with large spin $J$.
In this way, $\vev{p|\widehat{\cal O}_J(u\xi)|p}
\sim \vev{p|{\cal O}_J(0)|p} \sim (p\!\cdot\!n)^J \mu^{-2 {\mit\Gamma}_{\rm cusp}
(\alpha_s)\ln J}$ and
\be
\vev{p\,|\widetilde W_\Pi(\xi)|p} \e^{- i \xi (p \cdot n)}
\sim
\sum_{J} \frac{(-i\xi)^J}{J!} (p\!\cdot\!n)^J
J^{- 2 {\mit\Gamma}_{\rm cusp}(\alpha_s) \ln [(p \cdot n) \mu]}
\, .
\ee
For $\xi\gg 1$ the sum in the right-hand side of this relation receives the
dominant contribution from $J \sim - i \xi (p\!\cdot\!n)$ leading
to~\cite{KorMar93}
\be
\vev{p\,|\widetilde W_{\scriptstyle\Pi} (\xi)|p}
\sim
\vev{p|\widehat{\cal O}_{J}(0)|p} \bigg|_{J = - i \xi (p \cdot n)}
\sim \mu^{-2 {\mit\Gamma}_{\rm cusp} (\alpha_s) \ln [- i \xi (p \cdot n)]}
\, .
\label{W=O}
\ee
This relation establishes a correspondence between the Wilson line
in the adjoint representation and the matrix element of the
conformal operator analytically continued to large complex values
of the spin $J$. Notice that in the multi-color limit,
$N_c\to\infty$, one has $\vev{\widetilde W_{\scriptstyle\Pi} (\xi)
} = \vev{W_{\scriptstyle\Pi} (\xi)}^2$ with
$W_{\scriptstyle\Pi}(\xi)$ defined in the fundamental
representation.

\subsection{Multi-particle conformal operators}
\label{Section-Multiparticle}

\begin{figure}[t]
\unitlength1mm
\begin{center}
\mbox{
\begin{picture}(0,52)(67,0)
\put(10,0){\insertfig{12}{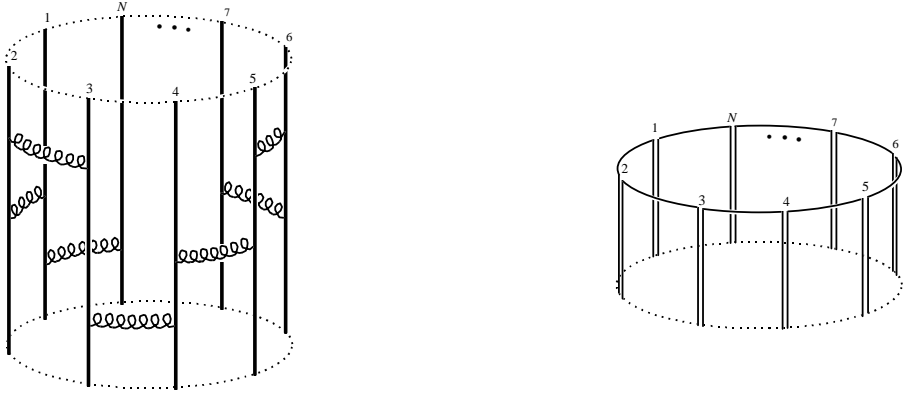}}
\end{picture}
}
\end{center}
\caption{\label{CylinderTopology} Leading Feynman diagram of cylinder topology
for a non-local $N$-particle operator in the large-$N_c$ limit in the light-cone
gauge when only nearest-neighbour interactions survive (left). The array of
$\scriptstyle\Pi$-shaped Wilson lines resulting from the figure on the left in
the Feynman gauge (right).}
\end{figure}

Let us generalize the above analysis to conformal operators built from three and
more primary fields. As before, we construct a nonlocal operator containing $N$
primary (quasi-partonic) fields on the light-cone and expand it in powers of
light-cone separations
\be
\tr
\bigg\{
X_1(0) {\mit\Phi}_n [0,\xi_2] X_2(\xi_2) \ldots X_N (\xi_N) {\mit\Phi}_{- n} [\xi_N,0]
\bigg\}
=
\sum_{j_2 \cdots j_N \ge 0}
\frac{(-i \xi_2)^{j_2}}{j_2!} \ldots \frac{(-i\xi_N)^{j_N}}{j_N!}
\mathcal{O}_{j_2\cdots j_N}(0)
\, .
\label{XX}
\ee
Here the Wilson lines run between two adjacent fields along the light-cone
to ensure the gauge invariance. The local composite operators have the form
\be
\mathcal{O}_{j_2\cdots j_N} (\xi)
=
\tr\bigg\{
X_1 (\xi) \left(i n\!\cdot\! {\cal D} \right)^{j_2}
X_2 (\xi) \ldots \left(i n\!\cdot\! {\cal D} \right)^{j_N}
X_N (\xi)
\bigg\}
\, ,
\label{O-N}
\ee
where, as before, the covariant derivative is defined in the adjoint
representation. The evolution of these operators under renormalization group
transformations is much more complicated as compared with the previous case due
to larger number of particles involved and complicated color flow. The latter can
be simplified by going over to the multi-color limit $N_c \to \infty$. In this
limit, in the axial gauge $n\!\cdot\!A (x) = 0$, the planar Feynman diagrams
contributing to
the left-hand side of \re{XX} have the topology of a cylinder%
\footnote{For $N = 3$ the result is exact for arbitrary $N_c$.} as shown in
Fig.\ \ref{CylinderTopology}.

In the multi-color limit, the operators \re{O-N} mix under renormalization among
themselves and with operators containing total derivatives. The mixing occurs
between the operators with the same total conformal spin $J+\sum_{k=1}^N j_{X_k}$
with $J=j_2+\ldots+ j_N$ and their number grows rapidly with $N$, even for the
forward matrix elements. In the latter case, the dimension of the mixing matrix
scales as $\sim J^{N-2}$ for large $J$. At $N=2$ this matrix has a single element
for arbitrary $J$. For the number of particles $N\ge 3$ the problem of
constructing multiplicatively renormalizable operators $\widehat O_J$ is reduced
to diagonalization of the mixing matrix whose size growes with $J$. As we will
argue below, the same problem is equivalent to solving a Schr\"odinger equation
for a Hamiltonian of a Heisenberg spin chain model. Before doing this, let us
analyze the picture from the point of view of the Wilson line formalism.

\subsubsection{Wilson line approach}
\label{WilsonApp}

Following the Wilson line approach \cite{KorMar93}, one can obtain the scaling
behavior of the anomalous dimensions of conformal operators at large spin $J$.
To this end, we examine Eq.\ \re{XX} for large light-cone separations $\xi_2\sim
\xi_3 \sim \ldots \sim \xi_N \gg 1$. As in the previous case, the nonlocal
light-cone operator in the left-hand side of \re{XX} is dominated by the
contribution of soft gluons with the wavelength $\sim \xi_k$. This allows one
to apply the eikonal approximation and replace the quantum fields by eikonal
phases, the Wilson lines in the adjoint representation evaluated along rays that
run along momenta of particles and terminate at the light-cone. Assuming for
simplicity that all particles have the same momentum $p_\mu$ we apply the eikonal
approximation%
\footnote{Here we used the relation between the eikonal phases defined in the
fundamental (${\mit\Phi}$) and adjoint ($\widetilde{\mit\Phi}$) representations,
$t^b[\widetilde {\mit\Phi}]_{ab} = {\mit\Phi}^\dagger t^a {\mit\Phi}$.}
\be
X_i (\xi)
\to
\e^{-i \xi (p \cdot n)} {\mit\Phi}_p^\dagger [-\infty,\xi]
\, b_i(p) \,
{\mit\Phi}_p [-\infty,\xi]
\, ,
\ee
where $b_i (p) = b_i^a (p) \, t^a$ is the annihilation operator of the $i$-th
particle and ${\mit\Phi}^\dagger_p [-\infty,\xi] = {\mit\Phi}_{- p} [\xi,
\infty]$. Then, the matrix element of the operator entering the left-hand side of
\re{XX} between the vacuum and the $N$-particle state $\tr\left\{b_1^\dagger (p)
b_2^\dagger (p) \ldots b_N^\dagger (p) \right\} \ket{0}$ is given in the
multi-color limit by (see Fig.\ \ref{CylinderTopology})
\be
\vev{\tr W_{\scriptstyle\Pi} [\xi_1,\xi_2]}
\vev{\tr W_{\scriptstyle\Pi} [\xi_2,\xi_3]}
\ldots
\vev{\tr W_{\scriptstyle\Pi} [\xi_N,\xi_1]}
\e^{- i (p \cdot n)(\xi_2 + \ldots + \xi_N)}
\, ,
\label{prod-W}
\ee
with $\xi_1 = 0$ and $|\xi_{k + 1} - \xi_k| \sim \xi \gg 1$. (In arriving at this
relation we applied the ``vacuum dominance'' property, $\vev{\tr W_1 \tr
W_2}=\vev{\tr W_1}\vev{\tr W_2}+O(1/N_c^2)$.) Here $W_{\scriptstyle\Pi} [\xi_2,
\xi_3]$ is the Wilson line in the fundamental representation evaluated along the
${\scriptstyle\Pi}$-like contour that runs along the momentum $p_\mu$ from $-
\infty$ to the point $\xi_2 n_\mu$, then along the light-cone to $\xi_3 n_\mu$
and returns to infinity along $-p_\mu$. Due to the presence of two cusps on the
integration contour, $\vev{\tr W_{\scriptstyle\Pi} [\xi_1, \xi_2]}$ acquires the
cusp anomalous dimension $\sim
\mu^{- {\mit\Gamma}_{\rm cusp}(\alpha_s) \ln [- i \xi (p \cdot n)]}$. In this
way, one finds from \re{prod-W} that the left-hand side of \re{XX} scales as
\be
\sim
\mu^{- N {\mit\Gamma}_{\rm cusp}(\alpha_s)
\ln [- i \xi (p \cdot n)]} \e^{- i (p \cdot n) (\xi_2 + \ldots + \xi_N)}
\, .
\ee
Let us now examine the scaling behavior of the right-hand side of \re{XX}. As
before, for large light-cone separations, $\xi\gg 1$, the sum is dominated by
the contribution of terms with $j_2 \sim \ldots \sim j_N \sim (-i \xi)$, or
equivalently $J = \sum_k j_k \sim - i \xi N$. The corresponding composite operator
$\mathcal{O}_{j_2\cdots j_N}(0)$ is renormalized multiplicatively and has the
anomalous dimension
\be
\gamma_J^{\rm (max)} = N {\mit\Gamma}_{\rm cusp}(\alpha_s) \ln J
\, .
\label{gamma-max}
\ee
We recall that this result was obtained in the multi-color limit
$N_c \to \infty$ for $J\gg 1$.

Anomalous dimensions of the $N-$particle conformal operators are defined as
eigenvalues of the mixing matrix. As we argue in the next section, they can be
parameterized by the set of $N-2$ nonnegative integers
$\ell=(\ell_1,\ldots,\ell_{N-2})$ such that $0\le \ell_1\le \ldots \le
\ell_{N-2} \le J$. Their total number equals the
size of the matrix and grows at large $J$ as $J^{N-2}$. For given
spin $J$, the possible values of the anomalous dimensions occupy
the band
\be
\gamma_J^{\rm (min)}
\le
\gamma_J \lr{\ell_1,\cdots \ell_{N-2}}
\le
\gamma_J^{\rm (max)} \, .
\label{bounds}
\ee
Eq.~\re{gamma-max} sets up the upper bound in the spectrum of the
anomalous dimensions.

We recall that \re{gamma-max} was obtained from the analysis of
the nonlocal operator in the left-hand side of \re{XX} at large
light-cone separations between the fields. To establish the lower
bound in \re{bounds}, one has to relax the latter condition by
allowing two or more fields to be closely located on the
light-cone. It is convenient to choose the axial gauge
$n\!\cdot\!A (x) = 0$ and consider
$\tr\left\{X_1(0)X_2(\xi_2)\ldots X_N(\xi_N)\right\}$ in the
region $\xi_2\ll 1$ and $|\xi_{k+1}-\xi_k|\sim \xi\gg 1$ with
$k\ge 2$. The fields $X_1(0)$ and $X_2(\xi_2)$ are separated along
the light-cone by a (relatively) short distance. They interact
with each other by exchanging particles with short wavelengths,
thus invalidating the eikonal approximation. At the same time, the
interaction of these two fields with the remaining fields still
occurs through soft gluon exchanges. Since the soft gluons with
the wavelength $\sim \xi$ can not resolve the fields $X_1(0)$ and
$X_2(\xi_2)$, they couple to their total color charge. This means
that one can replace the bilocal operator $X_1(0)X_2(\xi_2)$ by
its expansion over the two-particle conformal operators
$\widehat{\mathcal{O}}_j(0)$ and apply the eikonal approximation
to $\widehat{\mathcal{O}}_j(\xi_2)$ and remaining fields
$X_3(\xi_3),\ldots,X_N(\xi_N)$ afterwards. Repeating the analysis
one arrives at the same expression as \re{prod-W} with the only
difference that the factor $\vev{\tr W_{\scriptstyle\Pi} [\xi_1,
\xi_2]}$ is missing. As a consequence, the anomalous dimension of
the nonlocal light-cone operator is given at large $J$ by
\re{gamma-max} with $N$ replaced by $N-2$. In other words, the
coefficient in front of ${\mit\Gamma}_{\rm cusp} (\alpha_s)\ln J$
in Eq.~\re{gamma-max} counts the number of fields separated along
the light-cone by large distances $\xi \sim i J$.

This suggests that the minimal anomalous dimension in \re{bounds}
corresponds to the configuration when all fields in the left-hand
side of \re{XX} are grouped into two clusters on the light-cone
located at the points $0$ and $\xi \sim i J$, respectively,
leading to
\be
\gamma_J^{\rm (min)} = 2 {\mit\Gamma}_{\rm cusp}(\alpha_s) \ln J
\, ,
\label{gamma-min}
\ee
in agreement with the results of
Refs.~\cite{BraDerMan98,BraDerKorMan99,Bel99a,Bel99b,DerKorMan00,BKM}. In this
case, the right-hand side of \re{XX} receives contribution from the whole tower
of spin-$J$ conformal operators with the anomalous dimensions satisfying
\re{bounds}. Going over to the limit $\mu\to\infty$, one finds that the
right-hand side of \re{XX} receives dominant contribution from the operators with
minimal anomalous dimension given by \re{gamma-min}.

As a function of large spin $J$, the anomalous dimensions $\gamma_J
\lr{\ell_1,\cdots \ell_{N-2}}$ form the family of (non-intersecting)
trajectories labelled by the integers $\ell_1,\cdots\ell_{N-2}$. Since the width
of the band \re{bounds} scales as $\ln J$, while the total number of anomalous
dimensions $\gamma_J\lr{\ell_1,\cdots \ell_{N-2}}$ grows as a power of $J$, one
obtains that for $J\gg 1$ the distribution of the trajectories inside the band
can be described by a continuous function whose explicit form depends on the
coupling constant.  Going over from the weak to the strong coupling one finds
that the trajectories do not intercept as $\alpha_s$ increases although their
distribution inside the band \re{bounds} is modified.

\subsubsection{Hamiltonian approach to evolution equations}

The above analysis relies on the properties of Wilson loops. Let us now develop a
``dual'' picture based on the properties of conformal operators. In the
multi-color limit, the conformal operators are given by linear combinations of
composite operators \re{O-N} including the operators with total derivatives. As
was already discussed above, to construct $N$-particle conformal operators, one
has to decompose the tensor product of $N$ representations of the
$SL(2,\mathbb{R})$ group labelled by conformal spins, $j_k\equiv j_{X_k}$, of the
fields $X_i(\xi)$ and project out the nonlocal operator \re{XX} onto the spin-$J$
representation
\be
[j_1] \otimes [j_2] \otimes \ldots \otimes [j_n]
=
\sum_{J\ge 0} [J + j_1 + j_2 + \ldots + j_N]
\, .
\ee
Subsequently applying the rule for the sum of two $SL(2,\mathbb{R})$ spins,
$[j_1] \otimes [j_2] = \sum_{j_{12}\ge 0} [j_1 + j_2 + j_{12}]$, one finds
that the spin-$J$ representation has a nontrivial multiplicity $n_J =
(J+N-2)!/[J!(N-2)!]$. It is uniquely specified by the ``external'' conformal
spins $j_1,\ldots,j_N$ and ``internal'' spins $0\le j_{12} \le j_{123} \le
\ldots \le j_{12\ldots N-1}\le J$ with $j_{12\ldots k}$ defining the total
spin in the $(12\ldots k)$-channel. Each irreducible spin-$J$ component gives
rise to the following local composite operator
\be
\widehat{\mathcal{O}}_J^{\{j\}} (\xi)
=
P_J^{\{j\}} (i\partial_1,\ldots,i\partial_N)
\tr\left\{
X_1 (\xi_1) X_2 (\xi_2) \ldots X_N (\xi_N)
\right\}\bigg|_{\xi_1 = \ldots = \xi_N = \xi}
\, ,
\label{O-basis}
\ee
where $\{j\}\equiv (j_{12}, \ldots, j_{12\cdots N - 1})$ and $P_J^{\{j\}}
(x_1,\ldots,x_N)$ is a homogeneous polynomial of degree $N$ in momentum
fractions $x_k$. This polynomial is the highest weight vector of the
spin-$J$ $SL(2,\mathbb{R})$ representation of the discrete series. It
satisfies the system of differential equations
\ba
&&\lr{\vec L_{1}+\ldots+\vec L_{k}}^2 P_J^{\{j\}}
=
J_{12\cdots k} (J_{12\cdots k} - 1)P_J^{\{j\}}
\, , \qquad
(k = 2,\ldots,N)
\, ,
\nonumber
\\
&&\lr{L_{1}^++\ldots+L_{N}^+} P_J^{\{j\}} = 0\,,
\label{P-system}
\ea with
$J_{12\cdots k}=j_1+\ldots+j_k+ j_{12\cdots k}$ and $j_{12\cdots
N}\equiv J$. Here $\vec L_k=(L_k^+,L_k^-,L_k^0)$ are the
$SL(2,\mathbb{R})$ generators in the ``momentum'' representation
\be
L_k^- = - x_k
\, , \qquad
L_k^+ = 2 j_k \partial_{x_k} + x_k \partial_{x_k}^2
\, , \qquad
L_k^0 = j_k + x_k \partial_{x_k}
\, ,
\label{L's}
\ee
with $\vec L^2=(L^+L^-+L^-L^+)/2+L_0^2$. At $N = 2$ the solution to \re{P-system}
is given by Jacobi polynomials (see Eq.~\re{Conf-op}). For higher $N$, it can
be constructed iteratively as a product of the $N = 2$ solutions
\cite{BraDerKorMan99,Bel99a}.%
\footnote{One can obtain the same expression by subsequently applying the
fusion rules \re{OPE-conf} to the product of $N$ primary fields.}

The operators \re{O-basis} are transformed under the $SL(2,\mathbb{R})$ conformal
transformations as primary fields with the same conformal spin $J + \sum_k j_k$.
However, they do not have autonomous evolution and mix under renormalization with
each other. To construct multiplicative conformal operators, one has to
diagonalize the corresponding $n_J\times n_J$ mixing matrix. Its eigenstates
define the coefficients $c_{\ell,\{j\}}$ of the expansion of the conformal
operators over the basis \re{O-basis} and the corresponding eigenvalues provide
their anomalous dimensions
\be
\widehat{\mathcal{O}}_{J,\ell}(\xi)
=
\sum_{\{j\}} c_{\ell,\{j\}}\widehat{\mathcal{O}}_J^{\{j\}}(\xi)
\equiv
P_{J,\ell}(i\partial_1,\ldots,i\partial_N)
\tr\left\{
X_1 (\xi_1) X_2(\xi_2) \ldots X_N(\xi_N)
\right\} \bigg|_{\xi_1 = \ldots = \xi_N = \xi}
\, ,
\label{general-O}
\ee
where the subscript $\ell=(\ell_1,\ldots,\ell_{N-2})$ with $0\le \ell_1\le
\ldots\ell_{N-2}\le J$ enumerates different conformal operators, or equivalently homogenous
polynomials $P_{J,\ell} (x_1,
\ldots, x_N)=\sum_{\{j\}} c_{\ell,\{j\}} P_J^{\{j\}} (x_1,\ldots,x_N)$.

Thus, in distinction with the $N=2$ case, the conformal symmetry alone does not
allow one to construct multiplicatively renormalizable conformal operators for
$N \ge 3$. Nevertheless, it reduces the problem to diagonalizing the mixing
matrix of dimension $n_J$. This matrix has a number of remarkable properties.
To begin with, we notice that the homogenous polynomials entering \re{O-basis}
are orthogonal to each other with respect to the $SL(2,\mathbb{R})$ scalar product
\be
\vev{J,\{j\}|J',\{j'\}}
\equiv
\int_0^1 [d^N x]\, x_1^{2j_1-1} \ldots x_N^{2j_N-1}
P_J^{\{j\}} (x_1,\ldots,x_N) P_{J'}^{\{j'\}} (x_1,\ldots,x_N)
\sim
\delta_{JJ'}\delta_{jj'}
\, ,
\label{scal-prod}
\ee
where $[d^Nx] = dx_1 \ldots dx_N \delta (1 - \sum_kx_k)$ and integration goes
over $0\le x_k\le 1$. This follows from the fact that the $SL(2,\mathbb{R})$
generators \re{L's} are self-adjoint operators on the vector space endowed with
the scalar product \re{scal-prod}. Then, the mixing matrix can be interpreted as
a Hamiltonian acting on the Hilbert space \re{scal-prod}, $\vev{J,\{j\}|
\mathcal{H}_N|J',\{j'\}}$. Denoting $\ket{J,\ell}\equiv P_{J,\ell}(x_1,\ldots,x_N)$,
one can determine the anomalous dimensions of the conformal operators
$\gamma_J(\ell)$ as solutions to the $N$-particle Schr\"odinger equation
\be
\mathcal{H}_N\ket{J,\ell} = \gamma_J{(\ell)}\ket{J,\ell}
\label{Sch}
\ee
under additional (highest weight) condition for its eigenstates
\be
L_{\rm tot}^+ \ket{J,\ell} = 0
\, , \qquad
L_{\rm tot}^0 \ket{J,\ell} = (J + \sum_{k = 1}^N j_k) \ket{J,\ell}
\, ,
\ee
with $\vec L_{\rm tot}\equiv \sum_{k=1}^N \vec L_k$ being the total conformal
spin. The Hamiltonian $\mathcal{H}_N$ commutes with the total $SL(2,\mathbb{R})$
spin, $[\vec L_{\rm tot},\mathcal{H}_N]=0$ and
is a self-adjoint operator on the Hilbert space \re{scal-prod}. This ensures that
the anomalous dimensions $\gamma_J{(\ell)}$ take real values.

In perturbation theory, the Hamiltonian $\mathcal{H}_N$ can be obtained from
explicit calculation of Feynman diagrams in the multi-color limit $N_c \to
\infty$~\cite{BukFroKurLip85}. To the lowest order in $\alpha_s$, due to cylinder-like topology of the
planar diagrams, the interaction occurs only between nearest neighbors (see Fig.\
\ref{CylinderTopology})
\be
\mathcal{H}_N
=
\frac{\alpha_s N_c}{\pi}
\Big(
H_{12} + H_{23} + \ldots + H_{N1}
\Big)
+
{\cal O} \left( (\alpha_s N_c)^2 \right)
\, ,
\label{H_N}
\ee
where the two-particle Hamiltonian $H_{k,k+1}$ acts on the ``coordinates''
$x_k$ and $x_{k+1}$ only.%
\footnote{We recall that $x_k$ has the meaning of the momentum fraction
carried by the particle described by the field $X_k(\xi_k)$.} Conformal
invariance implies that $H_{k,k+1}$ depends on the sum of two $SL(2,\mathbb{R})$
spins
\ba
(\vec L_k + \vec L_{k+1})^2
\!\!\!&\equiv&\!\!\!
J_{k,k+1} (J_{k,k+1} - 1)
\nonumber
\\
&=&\!\!\!
-
x_1 x_2 ( \partial_1 - \partial_2 )^2
+
2(j_2 x_1 - j_1 x_2) (\partial_1 - \partial_2) + (j_1 + j_2)(j_1 + j_2 - 1)
\, ,
\label{Jk}
\ea
where $\partial_k = \partial/\partial x_k$. The general expression for
$H_{k,k+1}$ looks like
\be
H_{k,k+1} = \psi(J_{k,k+1}) + \ldots
\, ,
\label{H-2-particle}
\ee
where $\psi(x) = d\ln\mit{\Gamma}(x)/dx$ is the Euler psi-function. Here ellipses
denote additional, rational in $J_{k,k+1}$ terms which are subleading for
$J_{k,k+1}\gg 1$. In distinction with the first term in the right-hand side of
\re{H-2-particle}, they depend on the type of particles involved.

The two-particle Hamiltonian \re{H-2-particle} has a universal form for large
spins $J_{k,k+1} \gg 1$
\be
H_{k,k+1}=\ln J_{k,k+1} + \mathcal{O}\left((J_{k,k+1})^0\right)\,.
\label{H-2-log}
\ee
At $N = 2$ the Hamiltonian \re{H_N} equals $\mathcal{H}_{N=2} = 2 ({\alpha_s
N_c}/{\pi})H_{12}$ and its eigenvalues, $\gamma_{N = 2}(J) \sim 2 ({\alpha_s
N_c}/{\pi})\ln J$ for $J\gg 1$, define the anomalous dimension of the
two-particle conformal operator at weak coupling, Eq.~\re{gamma-as}. To find
the spectrum of anomalous dimension for $N \ge 3$, one has to solve the
Schr\"odinger equation \re{Sch} for the $N$-particle Hamiltonian \re{H_N}.
It turns out the quantum-mechanical system with the Hamiltonian \re{Sch}
possesses a hidden symmetry and is intrinsically related to Heisenberg
spin magnets.

We would like to emphasize that the equivalence between evolution equations and
dynamical Hamiltonian systems is a rather general phenomenon in Yang-Mills
theories. In particular, similar integrable structures appear in the Regge
asymptotics \cite{regge} with the evolution ``time" being the logarithm of the
energy scale $t_{\rm Regge} =
\ln s$ and in the low-energy behavior of the $\mathcal{N}=2$ effective action
with the ``time" $t_{\rm SYM}= \ln \Lambda_{\rm QCD}$ \cite{n=2}.

\subsubsection{Heisenberg spin chains}
\label{HeisenbergChain}

Let us substitute \re{H-2-particle} into \re{H_N} and define the following
Hamiltonian
\be
\mathcal{H}_N^{\rm quan}=
\frac{\alpha_s N_c}{\pi}\sum_{k=1}^N \psi(J_{k,k+1})\,,
\label{H-eff}
\ee
with $J_{N,N+1}\equiv J_{N,1}$. Notice that, in general, it differs from the
exact QCD Hamiltonian \re{H_N} by terms that vanish for $J_{k,k+1}\gg 1$.
Therefore one should expect that in spite of the fact that the fine structure of
the energy spectrum of two Hamiltonians may be different, their asymptotic
behavior for $J\gg 1$ is the same. Exact calculations at $N=3$ confirm such
expectations \cite{BraDerMan98,BraDerKorMan99,Bel99a,Bel99b,DerKorMan00,BKM}. The
main advantage of dealing with \re{H-eff} is that the Hamiltonian \re{H-eff}
possesses a set of integrals of motion, $q = (q_2, \ldots, q_N)$,
\be
[\mathcal{H}_N^{\rm quan}, q_n] = [q_n, q_m] = 0 \, ,
\ee
and, as a consequence, the Schr\"odinger equation for $\mathcal{H}_N^{\rm quan}$
turns out to be completely integrable
\cite{BraDerMan98,BraDerKorMan99,Bel99a,Bel99b,DerKorMan00,BKM}. The Hamiltonian
\re{H-eff} is well-known in the theory of integrable models. It has been
constructed in \cite{SklTakFad79,Fad80,FadTak79,KulResSkl81} as a generalization
of the celebrated spin-$1/2$ XXX Heisenberg magnet to higher spin representations
of the $SU(2)$ and $SL(2,\mathbb{R})$ groups.

Eq.~\re{H-eff} defines periodic Heisenberg spin chain model of length $N$ and
spins being the $SL (2,\mathbb{R})$ generators. The value of the spin at the
$k$-th site is given by the conformal spin of the corresponding primary field.
Such identification allows one to solve the spectral problem for the Hamiltonian
\re{H-eff} exactly by the Quantum Inverse Scattering Method
\cite{SklTakFad79,Fad80,FadTak79,KulResSkl81,TarTakFad83,Fad95}. In particular,
using the Lax operator for the XXX Heisenberg spin magnet
\be
\mathbb{L}_k(u)
=
\lr{
\begin{array}{cc}
u + i {L}^0_k & i {L}^+_k \\
i {L}^-_k     & u - i {L}^0_k
\end{array}
}
\label{Lax}
\ee
with $L_k^0$ and $L_k^\pm$ being the $SL(2,\mathbb{R})$ generators \re{L's}, one
can obtain the explicit form of the integrals of motion $q_2,\ldots, q_N$ from
the expansion of the transfer matrix $t_N(u)$ in powers of the spectral parameter
$u$
\be
t_N(u)
=
\tr
\left[ \mathbb{L}_1(u) \mathbb{L}_2(u)\ldots \mathbb{L}_N(u) \right]
=
2 u^N + q_2 u^{N - 2} + \ldots + q_N
\, ,
\label{t-N}
\ee
with $q_2 = - \vec L_{\rm tot}^2 + \sum_{k = 1}^N j_k (j_k - 1)$ depending on the
total spin of the system $J$. Due to complete integrability of the model, the
energy spectrum is uniquely specified by their eigenvalues, $E_N = E_N (q_2,
\ldots, q_N)$. Applying the Bethe Ansatz
\cite{SklTakFad79,Fad80,FadTak79,KulResSkl81,TarTakFad83,Fad95}, one can
calculate explicitly both the energy spectrum and the corresponding
eigenfunctions. We recall that the former defines the anomalous dimensions of
conformal operators, while the latter determine the polynomials $P_{J,\ell} (x_1,
\ldots, x_N)$ entering \re{general-O}.

For our purposes, we are interested in finding the large $J$ asymptotic
behavior of the energy spectrum. Assuming for simplicity that the
particles have the same $SL (2,\mathbb{R})$ spin, $j_1 = \ldots = j_N
\equiv j$ one obtains the following asymptotic expression for the energy
\cite{Kor95b,BraDerKorMan99,Bel99b}
\be
\gamma_{J}(\ell)
=
\frac{\alpha_s N_c}{\pi}
\left\{
\ln 2
+
\Re{\rm e}
\left[
\sum_{k = 1}^N \psi(j + i \lambda_k) + \gamma_{\rm E}
\right]
+
\mathcal{O}(1/J^{2N})
\right\}
\, ,
\label{gamma-int}
\ee
where $\lambda_k$ are roots of the transfer matrix \re{t-N}, $t_N(u) = 2
\prod_k (u-\lambda_k)$. According to \re{t-N}, the quantum numbers $q_k$ are
given by symmetric polynomials of degree $k$ in $\lambda_1,\ldots,\lambda_N$.
For $J\gg 1$ one finds that the roots are real and they scale differently with
$J$ in the upper and lower part of the spectrum. In the upper part of the
spectrum, all roots scale as $\lambda_k \sim J$ with $k = 1, \ldots, N$ leading
to $q_k = \mathcal{O}(J^k)$
\be
\gamma_J^{\rm (max)}
=
\frac{\alpha_s N_c}{\pi} \ln \left| \lambda_1\lambda_2\ldots \lambda_N \right|
=
\frac{\alpha_s N_c}{\pi} \ln q_N = N \frac{\alpha_s N_c}{\pi}\ln J \, .
\label{Gamma-up}
\ee
In the lower part of the spectrum, $\lambda_1 \sim -\lambda_2\sim \mathcal{O} (J)$
and $\lambda_k \sim J^0$ for $k\ge 3$ (notice that $\sum_k\lambda_k = 0$ due to
absence of the $u^{N-1}-$term in the right-hand side of \re{t-N}). This leads
to $q_k \sim J^2$ and
\be
\gamma_J^{\rm (min)}
=
\frac{\alpha_s N_c}{\pi} \ln \left| \lambda_1 \lambda_2 \right|
=
2 \frac{\alpha_s N_c}{\pi}\ln J
\, .
\label{Gamma-down}
\ee
We observe a perfect agreement of these expressions with Eqs.~\re{gamma-max} and
\re{gamma-min} obtained within the Wilson line approach.

In the Wilson line approach, the asymptotic behaviour of the anomalous
dimensions, Eqs.~\re{gamma-max} and \re{gamma-min} is tied to the scaling scaling
behaviour of the two-particle spins $J_{k,k+1}$ while in the Hamiltonian approach
Eqs.~\re{Gamma-up} and \re{Gamma-down} follow from similar behaviour of the roots
$\lambda_k$ of the transfer matrix \re{t-N}. Let us demonstrate that
$\lambda_k\sim J_{k,k+1}$ at large $J$.

By the definition, $J_{k,k+1}$ is the sum of two $SL(2,\mathbb{R})$ spins. Its
eigenvalues satisfy $j_k+j_{k+1} \le J_{k,k+1} \le J$, where $J$ is the total
spin of $N$ particles. Since $[\mathcal{H}_N,J_{k,k+1}]\neq 0$, one can not
assign a definite value of $J_{k,k+1}$ to the eigenstates of the Hamiltonian
$\mathcal{H}_N$. Nevertheless, for $J\to\infty$ the system of $N$ particles
approaches a quasiclassical regime in which quantum fluctuations are frozen and
the spins $J_{k,k+1}$ can be treated as classical variables. To see this one
notices that for quantum-mechanical systems defined by the Hamiltonian \re{H-eff}
the ``effective'' Planck constant equals unity, $\hbar=1$, and the energy scale
is defined by the total spin $J$. This suggests that for $J \gg
\hbar$ one can solve the Schr\"odinger equation
\re{Sch} by the WKB methods~\cite{Kor95b}.

In the WKB approach, one looks for the solution to \re{Sch} in the form
\be
\ket{J,\ell}\equiv P_{J,\ell}(x_1,\ldots,x_N) = \exp\lr{\frac{i}{\hbar}
S_0(\vec x)+ i S_1(\vec x) + \mathcal{O}(\hbar)}\,,
\label{WKB}
\ee
where $\vec x=(x_1,\ldots,x_N)$. To find the functions $S_0(\vec x)$, $S_1(\vec
x)$, $\ldots$, one requires that the wave function \re{WKB} has to be an
eigenstate of the transfer matrix \re{t-N}, or equivalently diagonalize
simultaneously the integrals of motion $q_2,\ldots,q_N$. In this way, one obtains
that the leading term $S_0(\vec x)$ satisfies the Hamilton-Jacobi equations in
the underlying classical system, while subleading terms can be  expressed in
terms of $S_0(\vec x)$. To go over to the classical limit, one applies the
operator of the two-particle spin defined in \re{Jk} to the WKB wave function
\re{WKB}
\be
\hbar^2(\vec L_k + \vec L_{k+1})^2 \e^{iS_0/\hbar}
=
\left\{
x_k x_{k+1} (p_k-p_{k+1})^2 + \mathcal{O}(\hbar)
\right\}
\e^{iS_0/\hbar}
\, ,
\label{WKB-L2}
\ee
where we have restored in the left-hand side the dependence on the Planck
constant and $p_k = - i \hbar\partial_{x_k} S_0(\vec x)$ is a classical momentum
of the $k$-th particle. In a similar manner, the $SL(2,\mathbb{R})$ spin
operators \re{L's} can be replaced by classical functions on the phase space
of $N$ particles
\be
L_{k,\, \rm cl}^- = -x_k
\, , \qquad
L_{k,\, \rm cl}^+ = - x_k p_k^2
\, ,\qquad
L_{k,\, \rm cl}^0 = i x_k p_k
\, .
\label{L-class}
\ee
Notice that the dependence on a single-particle spin disappears since
$j_k=\mathcal{O}(\hbar)$. One verifies that, in agreement with \re{WKB-L2},
\be
(\vec L_{k,\, \rm cl} + \vec L_{k+1,\, \rm cl})^2
=
(J_{k,k+1}^{\,\rm cl})^2
=
x_k x_{k+1} (p_k - p_{k+1})^2
\, .
\label{J-class}
\ee
Here $J_{k,k+1}^{\,\rm cl}$ defines the classical limit of the two-particle spin
$J_{k,k+1}$. Then, replacing $J_{k,k+1}\to \hbar J_{k,k+1}^{\,\rm cl}$ in
\re{H-eff}, one expands the Hamiltonian $\mathcal{H}_N^{\rm quan}$ in powers of
$\hbar$ and identifies the leading term of the expansion as the Hamiltonian of
the classical model
\be
\mathcal{H}_N^{\,\rm cl}
=
\frac{\alpha_s N_c}{2\pi}\sum_{k=1}^N
\ln \lr{J_{k,k+1}^{\,\rm cl}}^2
\, .
\label{H-class}
\ee
Remarkably enough, this Hamiltonian inherits integrability properties of
the quantum model. It contains a set of integrals of motion $q_k^{\rm cl}$
$(k = 2, \ldots, N)$
\be
\{ \mathcal{H}_N^{\,\rm cl}, q_k^{\,\rm cl} \}
=
\{ q_n^{\,\rm cl}, q_k^{\,\rm cl} \}
=
0
\, ,
\ee
with the Poisson bracket defined as
$\{f,g\}=\partial_{x_k}f\partial_{p_k}g-\partial_{p_k}f\partial_{x_k}g$.  To
obtain their explicit form, one replaces the $SL(2,\mathbb{R})$ spin operators
in \re{Lax} by their classical counterparts \re{L-class} and substitutes the
resulting expression for the Lax operator into \re{t-N}. This leads to
\be
q_n^{\,\rm cl}
=
\sum_{1\le j_1 < \cdots < j_n\le N} x_{j_1} \ldots x_{j_{n-1}}
x_{j_n} (p_{j_1}-p_{j_2})
\ldots (p_{j_{n-1}}-p_{j_n})(p_{j_n}-p_{j_1})\,,
\label{qn-class}
\ee
with $n=2,\ldots, N$. One observes that $\lr{q_N^{\,\rm cl}}^2 = \prod_{k=1}^N
\lr{J_{k,k+1}^{\,\rm cl}}^2$ and, therefore, the classical Hamiltonian \re{H-class}
can be written as
\be
\mathcal{H}_N^{\,\rm cl}
=
\frac{\alpha_s N_c}{\pi} \ln q_N^{\,\rm cl}
\, .
\label{H-asym}
\ee
By construction, $\mathcal{H}_N^{\,\rm cl}$ defines a classical limit of the
Heisenberg $SL(2,\mathbb{R})$ spin magnet model. Evaluating $\mathcal{H}_N^{\,\rm
cl}$ along the orbits of classical motion of $N$ particles, one obtains the
energy spectrum of the quantum magnet to the leading order of the WKB expansion,
or equivalently the large-$J$ behavior of the anomalous dimensions
$\gamma_J{(\ell)}$ of conformal operators, Eq.~\re{Sch}. According to
\re{H-asym}, this behavior is controlled by the large-$J$ scaling of the
``highest'' integral of motion $q_N^{\,\rm cl}$. In the WKB approach, the
spectrum of $q_N$ is obtained by imposing the Bohr-Sommerfeld quantization
conditions on the periodic orbits of classical motion of $N$
particles~\cite{Kor95b}. As was shown in Ref.\
\cite{BraDerKorMan99,Bel99a,Bel99b,DerKorMan00}, the WKB spectrum of anomalous
dimensions derived in this manner is in agreement with the exact expressions
\re{Gamma-up} and \re{Gamma-down}.

Finally, let us establish the relation between the roots $\lambda_k$ of the
transfer matrix and the two-particle spins $J_{k,k+1}^{\rm \,cl}$. Substituting
$t_N(u)=2\prod_k(u-\lambda_k)$ into \re{t-N} one obtains that the roots
parameterize the eigenvalues of the integrals of motion $q_n$. At large $J$,
replacing $q_n$ by their classical counterparts defined in \re{qn-class}, one
finds that $q_n^{\,\rm cl}$ are given by symmetric polynomials in
$\lambda_1,\ldots,\lambda_N$ of degree $n$ (with $n = 2,
\ldots, N$). Therefore, the large-$J$ behavior of the momenta $p_k
- p_{k+1}$ and the roots $\lambda_k$ are in one-to-one correspondence with each
other. In particular, in the upper part of the spectrum, Eq.~\re{Gamma-up}, one
gets from $\lambda_k\sim J$ that $p_k-p_{k+1}\sim J$, or equivalently
$J_{k,k+1}^{\rm\,cl}\sim J$ for $k=1,\ldots,N$. In similar manner, in the lower
part of the spectrum, Eq.~\re{Gamma-down}, $\lambda_k\sim J^0$ leads to
$p_k-p_{k+1}\sim J^0$ and $J_{k,k+1}^{\rm\,cl}\sim J^0$. We recall that the
$x_k$-variables  entering \re{J-class} have the meaning of momentum fractions
carried by $N$ particles described by quantum fields $X_k(\xi_k)$. Then,
conjugated to them the $p_k$-variables are the light-cone coordinates of the same
fields, $p_k\equiv \xi_k$. Thus, in the upper and the lower part of the spectrum
one has $\xi_k-\xi_{k+1}\sim J$ and $\xi_k-\xi_{k+1}\sim J^0$, respectively.
These properties are in agreement with the results obtained in section
\ref{WilsonApp} within the Wilson line approach.

We notice that the anomalous dimensions of $N-$particle conformal operators,
Eq.~\re{Gamma-up} and \re{Gamma-down}, were obtained using the lowest-order
expression for the QCD evolution kernels whereas Eqs.\ \re{gamma-max} and
\re{gamma-min} hold to all orders in the coupling constant. The above analysis
suggests that at large $J$ the higher order corrections modify the classical
Hamiltonian \re{H-class} and \re{H-asym} in the following way
\be
\mathcal{H}_N^{\,\rm cl}
=
\frac{1}{2} {\mit\Gamma}_{\rm cusp}(\alpha_s)
\sum_{k = 1}^N \ln \lr{J_{k,k+1}^{\,\rm cl}}^2
=
{\mit\Gamma}_{\rm cusp}(\alpha_s)\ln q_N^{\,\rm cl}
\, ,
\ee
with $J_{k,k+1}^{\,\rm cl}$ and $q_N^{\,\rm cl}$ given by the same expressions
as before, Eqs.~\re{J-class} and \re{qn-class}, respectively.

\section{Cusp anomaly at weak coupling}
\label{WeakCusp}

Let us revisit the computation of the cusp anomalous dimension to the lowest
order of perturbation theory aiming on an analogy with the stringy computation
of Wilson loops within the AdS/CFT framework. As we will see momentarily, the
cusp anomaly in the weak coupling regime can be interpreted as a {\sl quantum\/}
transition amplitude for a test particle propagating in the radial time $\ln r$
and the angular coordinate $\theta$. This should be compared with the strong
coupling calculation~\cite{DruGroOog99,Kru02,Mak02}, in which the same quantity
is given by a {\sl classical\/} action function for a particle propagating on
the same phase space.

\subsection{Cusp anomaly in perturbation theory}

\begin{figure}[t]
\unitlength1mm
\begin{center}
\mbox{
\begin{picture}(0,42)(60,0)
\put(10,0){\insertfig{10}{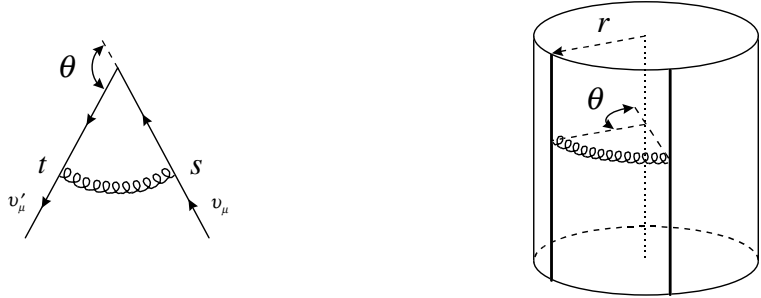}}
\end{picture}
}
\end{center}
\caption{\label{Cylinder} Cusp anomaly in one-loop approximation (left) and
its interpretation in radial quantization formalism (right).}
\end{figure}

To the lowest order in the coupling constant, the Wilson line expectation value
is given by
\be
W = \vev{P \exp\lr{i \oint_{\wedge} dx_\mu A^\mu(x)}}= 1 + \frac{(ig)^2}{2} t^a
t^a
\int_C dx_\mu\int_C dy_\nu \, D^{\mu\nu}(x - y) + \mathcal{O}(g^4)
\, ,
\label{W-1-loop}
\ee
where $D_{\mu\nu}(x - y) \delta^{ab} = \vev{0|T A^a_\mu(x)A^b_\nu(y)|0}$ is a
gluon propagator and $t^at^a=N_c$ is the Casimir operator in the adjoint
representation of the $SU(N_c)$. To calculate the cusp anomaly we choose the
integration contour $C$, see Fig.\ \ref{Cylinder} (left), to be the same as for
the Isgur-Wise form factor, Eq.~\re{W}. In this way, we obtain
\be
W (v\!\cdot\!v')
=
1 - \frac{\alpha_s N_c}{\pi} \Big( w (v\!\cdot\!v') - w(1) \Big)
+
\mathcal{O}(\alpha_s^2)
\, ,
\label{one-loop}
\ee
where $v_\mu$ and $v'_\mu$ are tangents to the integration contour in the
vicinity of the cusp, $v^2 = v'{}^2 = 1$, $v\!\cdot\! v' = \cosh\theta$,
$w(1) = w(v\!\cdot\!v) = w(v'\!\cdot\!v')$ and
\be
w (v\!\cdot\!v')
=
\int^0_{-\infty} d s \int_0^\infty d t \frac{v \!\cdot\! v'}{( v s - v' t )^2}
\, ,
\label{w}
\ee
with $s$ and $t$ being proper times. Going over to higher orders in $\alpha_s$,
one takes into account that the Wilson loop possesses the property of non-abelian
exponentiation~\cite{Ste81}
\begin{equation}
\label{WilsonExponentiation}
\ln W = \sum_k \lr{ \frac{\alpha_s N_c}{\pi} }^k w_k \,,
\end{equation}
where the weights $w_k$ receive contribution from diagrams to the $k$-th order
in $\alpha_s$ with maximally nonabelian color structure. In our case, the
exponentiation property states that $w_1=w(v\cdot v')-w(1)$.

It is straightforward to perform the integration in \re{w}. For our purposes,
however, we change the integration variables to $r_> = \max(t, - s)$ and
$r_< = \min(t, - s)$ and apply the identity
\be
\frac{v \!\cdot\! v'}{( v s - v' t )^2}
=
- \frac1{r_>^2}
\left\{
\sum_{n = 1}^\infty
(-1)^n \lr{\frac{r_<}{r_>}}^n \frac{(n + 1)^2}{n(n + 2)} \, U_n (v\!\cdot\! v')
+
\frac{1}{4}
\right\}
\, ,
\label{Chebyshev}
\ee
where $U_n(\cosh\theta)=\sinh((n + 1)\theta)/\sinh \theta$ are Chebyshev
polynomials of the second kind \cite{BatErd53-2}. Its substitution into
\re{w} leads to
\be
\label{WinTermsOfChebyshev}
w(v \!\cdot\! v')
=
- 2
\left\{
\sum_{n = 1}^\infty (-1)^n \frac{n + 1}{n(n + 2)} \, U_n (\cosh \theta)
+
\frac{1}{4}
\right\}
\,
\int_{r_{\rm min}}^{r_{\rm max}}\frac{dr_>}{r_>}
=
\theta\coth\theta \, \ln(\mu r_{\rm max})
\, ,
\ee
where $r_{\rm min}\sim 1/\mu$ and $r_{\rm max}$ are ultraviolet and infrared
cut-offs, respectively. Combining together \re{one-loop} and
\re{WinTermsOfChebyshev}, one verifies that the Wilson line satisfies the
evolution equation \re{RG}.

Eqs.~\re{Chebyshev} and \re{WinTermsOfChebyshev} have a simple interpretation
within the radial quantization approach \cite{FubHanJac73,Gal84}. In this
formalism, one performs a quantization procedure using four-dimensional
(Euclidean) polar coordinates $r^2=x_\mu^2$ and $v_\mu = x_\mu/r$. This allows
one to separate the dynamics in the radial and angular coordinates by decomposing
the propagators of fields over partial waves defined as eigenstates of the
operator of the angular momentum
\be
{\bit L}^2 = \frac18 l^{\mu\nu}l_{\mu\nu} \, , \qquad l_{\mu\nu} =
i \lr{x_\mu\partial_\nu-x_\nu\partial_\mu} \, ,
\ee
which has the meaning of the Laplace-Beltrami operator on a sphere $v_\mu^2 = 1$.
Then, in the radial quantization the lowest order contribution to the cusp
anomaly, Eq.~\re{WinTermsOfChebyshev}, takes the factorized form~\cite{Gal84}
\begin{equation}
w(v \!\cdot\! v')
= - \frac{1}{2}
\Big( \langle -v' | 1/\bit{L}^2 | v \rangle + 1 \Big) \int {d \ln r}
\, ,
\end{equation}
where $\ket{v}$ denotes a point on the (hyper)sphere $SO(3,1)/SO(3)$ defined by
the unit vector $v_\mu$ and additional minus sign inside $\bra{-v'}$ takes into
account that two tangents have opposite orientation at the cusp. Thus at the
weak coupling, the cusp anomalous dimension is given by
\be
{\mit\Gamma}_{\rm cusp} (\theta; \alpha_s)
=
- \frac{\alpha_s N_c}{2 \pi}
\Big(
\langle - v' | 1/\bit{L}^2 | v \rangle
-
\langle v| 1/\bit{L}^2 | v \rangle
\Big)
+
\mathcal{O}(\alpha_s^2)
\,.
\ee
Let us demonstrate that the matrix elements entering this expression coincide
with the propagator of a test particle on the time-like hyperboloid
$v_\mu^2 = v_0^2 - v_1^2 - v_2^2 - v_3^2 = 1$.

\subsection{Particle propagation on a sphere}
\label{PropagationSphere}

To avoid complications due to the infinite volume of the (noncompact)
$SL(2,\mathbb{R})$ group manifold, we will calculate the propagator of a particle
on a unit sphere $S^3$ and perform an analytic continuation from Euclidean to
Minkowski signatures afterwards.

\begin{figure}[t]
\unitlength1mm
\begin{center}
\mbox{
\begin{picture}(0,41)(92,0)
\put(30,0){\insertfig{12}{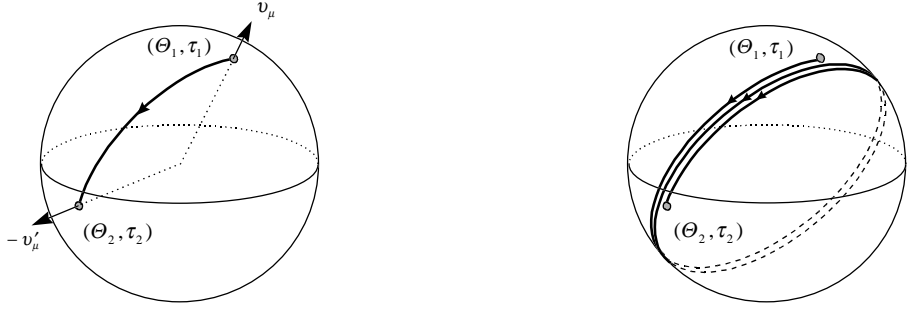}}
\end{picture}
}
\end{center}
\caption{\label{TopPropagator} Transition amplitude on a sphere between two
points $v_\mu$ and $- v'_\mu$ with $\theta_2-\theta_1=\pi-\theta$ and
$\tau_2-\tau_1=\tau$ (left). Classical trajectories which saturate the amplitude
are multiple windings around the sphere over principles circles, e.g., $\ell = 2$
trajectories (right), separated from each other for illustration purposes only. }
\end{figure}

The transition amplitude for the particle on the $S^3$ sphere to go from the
point $v_\mu$ to $- v'_\mu$ is equal to the sum over all paths $P$ connecting
these two points, Fig.\ \ref{TopPropagator},
\be
G[v,-v'] = \sum_{P \in S^3} \e^{-iA[P]}
\, ,
\label{part-fun}
\ee
where summation goes over connected paths $P$ of the lenght $A[P]$ on the
$S^3$ sphere. With an arbitrary point on the sphere, $v_\mu \in S^3$, one
can associate an element of the $SU(2)$ group
\be
g_v = v_0 + i\sum_{a = 1}^3 v_a \sigma_a
\, , \qquad
\tr \, [ g^{-1}_{v_1} g^{\phantom{1}}_{v_2} ] = 2 (v_1\!\cdot\! v_2)
\, ,
\label{g_n}
\ee
with $\sigma_a$ being Pauli matrices. Eq.~\re{part-fun} defines a quantum
dynamics of a particle on the $SU(2)$ group manifold. Introducing local
(angular) coordinates on the sphere $X^a = (\rho,\tau,\phi)$
\be
v_0 = \cos\rho\cos\tau
\,,\quad
v_1 = \sin\rho\sin\phi
\,,\quad
v_2 = \sin\rho\cos\phi
\,,\quad
v_3 = \cos\rho\sin\tau
\,,
\ee
one obtains the metric on this manifold as
\be
ds^2
=
\frac{1}{2} \tr \left[ g_v^{-1} d g^{\phantom{1}}_v \right]^2
=
- d \rho^2 - \sin^2 \rho \, d \phi^2 - \cos^2 \rho \, d \tau^2
\equiv
G_{MN}(X) dX^M dX^N
\, .
\label{metric}
\ee
The action of a particle has the meaning of the length of the path on this
manifold
\be
A[P]
=
\int_0^1 d\sigma \sqrt{- G_{MN}(X) \partial_\sigma X^M \partial_\sigma X^N}
=
\int_0^1 d\sigma
\left\{
\frac{e(\sigma)}{4}
-
\frac{1}{e(\sigma)} {G_{MN}(X) \partial_\sigma X^M \partial_\sigma X^N}
\right\}
\, ,
\label{action}
\ee
where $\sigma$ is a local coordinate on the trajectory. The two expressions
coincide upon extremizing with respect to the einbein field $e (\sigma)$.

Following, for instance,  \cite{Pol-book}, one can express this path integral as an integral over
the proper time $\tau = \int_0^1 \, d\sigma e(\sigma)$ of the matrix element of
the transition operator
\be
G[v, - v']
=
- i \int_0^\infty d \tau \vev{-v'| \e^{i\tau {\cal H}} |v}
=
\langle -v' |\, 1/{\cal H}\, | v \rangle
\, .
\label{G}
\ee
Here the Hamiltonian ${\cal H} = {\bit L^2}$ is the Laplace-Beltrami operator
on the $S^3$. It defines a quantum-mechanical model (symmetric top) with the
$SU(2)$ dynamical symmetry. Since the Hamiltonian coincides with the Casimir
operator of this group, its eigenstates correspond to unitary irreducible
$SU(2)$ representations of spin $j = 0, \frac12, 1, \ldots$
\be
\mathcal{H} \,Y^{j}_{m_1 m_2} (g)
=
j(j + 1) Y^{j}_{m_1m_2} (g)
\, , \qquad
Y^{j}_{m_1m_2} (g)
=
(2j + 1)^{1/2} d^j_{m_1m_2} (g)
\,,
\ee
where $d^j_{m_1m_2}(g) = \vev{j,m_1|d^j(g)|j,m_2}$, the matrix elements of the
spin-$j$ group representation $- j \le m_1, m_2 \le j$, are the well-known
Wigner functions \cite{Vil68}. The transition matrix element can be expanded
over characters of the $SU(2)$ representation
\ba
\vev{- v'| \e^{i\tau {\cal H}} |v}
\!\!&=&\!\!
\sum_{j = 0, \frac12, 1, \ldots}
(2j + 1)
\tr \left[ d^j (g_{-v'}^{-1}) d^j (g^{\phantom{1}}_v) \right]
\e^{i \tau j(j + 1)}
\nonumber\\
&=&\!\!
\sum_{j = 0, \frac12, 1, \ldots} (2j + 1)
\chi_j [g_{-v'}^{-1} g_v^{\phantom{1}}] \e^{i \tau j(j + 1)}
\, .
\label{vev-1}
\ea
Here, the $SU(2)$ character is defined as \cite{Vil68}
\be
\chi_j [g_v]
=
\sum_{m = - j}^j d^j_{mm} (g_v)
=
\frac{\sin (2j + 1){\mit\Theta}/2}{\sin{\mit\Theta}/2}
\, ,
\label{chi}
\ee
where the Euler angle is defined as $\cos({\mit\Theta}/{2}) = \tr g_v/2$ with
$g_v$ given by \re{g_n} leading to ${\mit\Theta} = 2(\pi-\theta)$. Substituting
\re{chi} into \re{vev-1} and making use of the second relation in \re{g_n}, one
calculates the matrix element entering \re{G} as
\be
\vev{-v'| \e^{i\tau \bit{\scriptstyle L}^2} |v}
=
\sum_{j = 0, \frac{1}{2}, 1,\ldots}
(2j + 1) \e^{i j(j + 1) \tau}
\frac{\sin((2j + 1) \tilde\theta)}{\sin \tilde\theta}
\, ,
\label{theta}
\ee
where $\tilde \theta = \pi - \theta$ is the angle between the vectors $v_\mu$
and $- v'_\mu$. Substituting \re{theta} into \re{G}, one notices that the
$j = 0$ term in \re{theta} leads to a divergent contribution upon integration
in the right-hand side of \re{G}. It does not depend, however, on the cusp
angle $\theta$ and cancels in the difference $G[v,-v'] - G[v,v]$ leading to
\be
{\mit\Gamma}_{\rm cusp} (\theta; \alpha_s)
=
- \frac{\alpha_s N_c}{2 \pi} \Big( G[v, - v'] - G[v, v] \Big)
=
\frac{\alpha_s N_c}{\pi} \lr{\theta\cot\theta - 1}
\, ,
\label{G0}
\ee
where $\theta$ is the Euclidean cusp angle $(v\!\cdot\!v') = \cos \theta$.
Going over to Minkowski space, we continue $\theta\to i\theta$ and reproduce
the correct expression for the cusp anomalous dimension, Eq.~\re{GammaCusp},
upon identification of the Casimirs $C_F = N_c$.

The following comments are in order. The sum in \re{theta} can be
expressed, via the Poisson summation formula, as a derivative of Jacobi
theta functions, namely,
\begin{eqnarray*}
\vev{-v'| q^{\bit{\scriptstyle L}^2} |v}
=
\frac{1}{2 q^{1/4} \sin\theta}
\frac{\partial}{\partial \tilde\theta}
\left[
\theta_2 ( \tilde\theta/\pi, q )
+
\theta_3 ( \tilde\theta/\pi, q )
\right]
\, ,
\end{eqnarray*}
where $q=\e^{i\tau}$ and $\tilde\theta=\pi-\theta$.
One can rewrite the same expression as
\ba
\vev{-v'| \e^{i\tau \bit{\scriptstyle L}^2}|v}
\!\!\!&=&\!\!\!
- 2 \frac{1}{\sin\theta} (-i \pi \tau)^{-3/2}
\sum_{\ell = -\infty}^\infty
(\tilde \theta + 2 \ell \pi)
\e^{-i (\tilde\theta + 2 \ell \pi)^2/\tau - i \tau/4}
\nonumber
\\
&=&\!\!\! \frac{1}{\pi\sin\theta} (-i \pi \tau)^{-1/2}
\frac{\partial}{\partial {\tilde\theta}}
\sum_{\ell = - \infty}^\infty
\e^{-i (\tilde\theta + 2 \ell \pi)^2/\tau - i \tau/4}
\, .
\label{vev-dual}
\ea
Eq.~\re{vev-dual} has a remarkably simple physical meaning.
As was shown in \cite{Sch67,Dow71,MarTer79,MenOno81}, see also Refs.\
\cite{Dur84,BohJun86}, the semiclassical expression for the transition
amplitude on the $SU(2)$ group manifold coincides with the exact solution,
Eq. \re{vev-dual}, i.e.\ the path integral collapses from a sum over all
paths to a sum over classical paths. In the semiclassical approach, the
right-hand side of \re{vev-dual} comes about as a sum over classical
trajectories ``dressed'' by quadratic fluctuations. The classical trajectories
are geodesics and run along the principal circle on the unit sphere between
the two points, $v_\mu$ and $-v'_\mu$, and wrap around this circle $\ell$-times
in the (anti-)clockwise direction depending on the sign of $\ell$. The
trajectories fall into two homotopy classes depending on the $\ell$-parity.
Denoting $\theta(\sigma)$ the angular variable on this circle, one can
parameterize classical trajectories as $\theta(\sigma) = ( \tilde\theta +
2 \pi \ell)\, \sigma/\tau$ with $0\le \sigma\le \tau$. The metric \re{metric}
on the classical trajectories equals
\be
ds^2 = G_{MN} dX^M dX^N = -d\theta^2
\ee
and the classical action \re{action} is given in the gauge $e(\sigma) =
{\rm const}$ by
\be
A_{\rm cl}[P]
=
\int_0^\tau d\sigma \lr{\frac1{4} + \dot\theta(\sigma)^2}
=
\frac{\tau}4 + \frac1{\tau}(\tilde\theta + 2 \pi \ell)^2
\, .
\ee
Coming back to the original sum \re{part-fun}, we conclude that
the exact expression for the transition amplitude \re{vev-dual},
and as a consequence the cusp anomalous dimension \re{G0}, is
given by the sum over classical trajectories. This property of the
path integral is a manifestation of the Duistermaat-Heckman
``localization'' phenomenon in quantum dynamical systems on Lie
groups \cite{DuiHec82}. 

The derivation of \re{theta} was based on the identification of the unit sphere
$S^3$ as the $SU(2)$ group manifold. Going over from Euclidean to Minkowski
kinematics, one has to substitute the sphere $S^3$ by the time-like hyperboloid
$H_3^+$, or equivalently the $3$-dimensional Lobachevsky space AdS${}_3$. The
appropriate group manifold is provided by the $SO(3,1)/SO(3)$ coset. In
distinction with the previous case, the dynamical symmetry group is noncompact
and we have to deal with quantum mechanics on the space of constant negative
curvature. The analysis goes along the same line as above with the only
difference that the $SU(2)$ representations of spin-$j$ have to be substituted
by the unitary, continuous representations of the $SO(3,1)$ group:
\begin{itemize}
\item fundamental series: $j = - 1/2 + i \nu/2$ with $- \infty < \nu < \infty$
\item complementary series: $- 1 < j < 0$
\end{itemize}
It turns out that the resulting expression can be obtained from \re{theta}
through analytical continuation in spin $j$. To see this, one applies the
Barnes-Mellin transformation to rewrite the sum over half-integer $j$ in
\re{theta} as a contour integral over the complex spin that runs parallel
to the imaginary axis to the right from $j = 0$
\be
G[v, - v'] - G[v,v]
=
- \int_{\delta - i \infty}^{\delta + i \infty} \frac{dj}{2\pi i}
\frac{\pi}{\sin(2\pi j)}
\frac{(2j + 1)^2}{j(j+  1)}
\left\{
\frac{\sin((2j + 1)\theta)}{(2j + 1)\sin\theta} - 1
\right\}
\,.
\label{G1}
\ee
with $0 < \delta < 1$. One verifies that moving the integration contour to the
right and picking up the residues at half-integer $j$ one reproduces the known
expression for the $SU(2)$ propagator, Eq.~\re{G0}. Let us now move the
integration contour in \re{G1} to the left parallel to the imaginary axis until
it reaches $\Re{\rm e} j = - 1/2$. Since the integrand has a pole at $j = 0$, the
deformed contour will contain an additional addendum that encircles the segment
$-1/2 < j < 0$ on the real axis. Changing the integration variable as $j = - 1/2
+ i \nu/2$ and going over to Minkowski kinematics, $\theta \to i\theta$, one
finds
\be
G[v, - v'] - G[v, v]
=
\lr{\int_{-\infty}^\infty + \, \frac12 \oint_{\gamma}}
\frac{d\nu\,\nu}{2(1 + \nu^2)}
\frac{\sin(\nu\,\theta)}{\sinh(\pi\nu)\sinh\theta}
\, .
\label{G2}
\ee
Here in the second integral we made use of the symmetry of the integrand under $j
\to -1-j$ and extended the integration to the contour $\gamma$, which encircles
the segment $[-i,i]$ in the anticlockwise direction. The two integrals in the
right-hand side of \re{G2} describe the contribution of the fundamental and
complimentary series, correspondingly. Since the integrand in \re{G2} is an odd
function of $\nu$, the former integral vanishes, while the latter is given by the
residue at the poles $\nu = \pm i$, or equivalently $j = 0,-1$. The resulting
expression for the propagator \re{G2} coincides with \re{G0} upon replacing
$\theta\to i\theta$. Similar to the previous case, one can expand the cusp
anomalous dimension as the sum over classical trajectories on the hyperplane
$v_0^2 - v_1^2 = 1$ defined by the time-like vectors $v_\mu$ and $v'_\mu$ in the
AdS space $v_0^2 - v_1^2 - v_2^2 - v_3^2 = 1$. Since the trajectories do not
``penetrate'' into transverse $(n_2,n_3)$-directions, the sum will be the same if
one changes the metric of the AdS${}_3$ space from Euclidean to Lorentzian
signature, $v_3 \to i v_3$. It is this version of the AdS${}_3$ space that one
encounters in the strong-coupling calculation of the cusp anomaly
\cite{GubKlePol02,Kru02,Mak02}.

\subsection{Analytic structure of cusp anomaly}

Examining \re{G0} one faces a paradox. By definition, $\cos\theta = (v\!\cdot\!
v')$, the cusp angle is defined up to $\theta \to \theta + 2 \pi n$ with $n$ an
arbitrary integer. At the same, \re{G0} is not invariant under this
transformation indicating that ${\mit\Gamma}_{\rm cusp} (\theta; \alpha_s)$ is a
{\sl multivalued} function of the cusp angle. To understand the origin of this
non-analyticity we observe that the sum in \re{theta} diverges at $\theta \to
\pi$, or equivalently $v'_\mu = - v_\mu$. This divergence has a simple meaning in
terms of the sum over random paths on the $S^3$-sphere, Eq.~\re{part-fun}. For $0
\le \theta < \pi$ the minimal length path connecting the points $v_\mu$ and
$v'_\mu$ is unique. For $\theta = \pi$ the points $v_\mu$ and $v'_\mu$ are
opposite poles on the sphere, the length of the minimal path equals $\pi$ and the
number of such paths is infinite. The same singularity has a clear meaning in QCD
in context of the heavy quark form factor. We recall that $v_\mu$ and $v'_\mu$
are velocities of the heavy quark before and after interaction with space-like
external momentum $q$, $- q^2 \equiv Q^2 = - m^2 (v - v')^2 = 2 m^2 [
(v\!\cdot\!v') - 1 ] > 0$. Re-expressing the cusp anomalous dimension
\re{GammaCusp} as a function of $x \equiv 4 m^2/Q^2=-1/\sin^{2}(\theta/2)$,
\begin{equation}
{\mit\Gamma}_{\rm cusp} (\theta; \alpha_s) = - \frac{\alpha_s N_c}{\pi}
\frac{1}{\sqrt{1 + x}}
\left\{
\sqrt{1 + x}
-
\left(1 + \frac{x}{2} \right) \ln \frac{\sqrt{1 + x} + 1}{\sqrt{1 + x} - 1}
\right\}
\, ,
\label{Cusp-x}
\end{equation}
one finds that the singularity at $x = - 1$, i.e.\ in the non-physical point
$Q^2 = - 4 m^2$, corresponding to $\theta = \pi$. The heavy quark form factor,
analytically continued from $Q^2 > 0$ to $Q^2 < 0$, describes the threshold
creation of a pair of heavy quarks. For $Q^2 > 4m^2$, i.e.\ above the threshold,
the cusp anomalous dimension acquires an imaginary part.

\subsection{Cusp anomalous dimension and 2D gauge theory}

In the previous section, we identified the one-loop cusp anomaly with a
transition amplitude for a test particle on the $SL (2,\mathbb{R})$ group
manifold. As a next step, we will express this amplitude in terms of a
disk partition function in a two-dimensional gauge theory and use
the stringy representation of the latter. We will demonstrate that the
emerging stringy description of the cusp anomaly at weak coupling is
different from the one dictated by the Nambu-Goto string.

In two dimensions, the Yang-Mills theory does not contain transverse gauge
degrees of freedom and, therefore, it can be reduced to a quantum-mechanical
model. Its partition function on an arbitrary Euclidean two-dimensional
manifold $\Sigma$ of genus $G$ with the metric tensor $g_{\mu\nu}$ can be
calculated through the heat kernel expansion \cite{Mig75}
\be
\label{PartitionFunction}
{\cal Z} [g^2 {\cal A}]
=
\int \mathcal{D}A_\mu \,
\e^{
- \frac{1}{g^2} \int_{\Sigma} d^2x \sqrt {\det g_{\mu\nu}} \tr F^2
}
=
\sum_{R} (\dim R)^{2 - 2G} \e^{- g^2 {\cal A} \, C_2(R)/2}
\, ,
\label{Z-2D}
\ee
where $F \equiv F^a_{01}[A] t^a$ is the only nontrivial component
of the strength tensor $F^a_{\mu\nu} [A]$ with generators in the
fundamental representation normalized as $\tr \, (t^a t^b) =
\delta^{ab}/2$. ${\cal A}$ is the area of the target manifold
$\Sigma$. The sum in the right-hand side of
(\ref{PartitionFunction}) runs over the unitary representations
$R$ of the gauge group of dimension $\dim R$ and quadratic Casimir
$C_2(R)$. As in section \ref{PropagationSphere}, we will consider
the $SU(2)$ gauge group and perform analytical continuation to the
$SL (2, \mathbb{R})$ group afterwards. In that case, $C_2(R) = j(j
+ 1)$ and $\dim R = (2j + 1)$ with $j$ being non-negative
(half)integer. Then, the sum in \re{Z-2D} has a striking
resemblance with a similar sum defining the transition amplitude
\re{vev-1}.

To make the correspondence exact, one introduces the amplitude of the
two-dimensional Yang-Mills theory on a disk, with radial coordinate $x^0$,
$0 \leq x^0 \leq T$, and angular $x^1$, $0 \leq x^1 \leq L$, of area
${\cal A} = LT/2$, and a holonomy at its boundary $C = \partial \Sigma$,
\begin{eqnarray*}
U = P \exp \lr{i \oint_C d x \cdot A(x)}
\, .
\end{eqnarray*}
The partition function of the disk is \cite{Mig75,Rus90}
\begin{equation}
\label{DiskAmplitude}
{\cal Z} [U; g^2 {\cal A}]
=
\int \mathcal{D}A_\mu \,
\delta \left( P \e^{i \oint_C dx \cdot A (x)},U \right)
\e^{
- \frac{1}{g^2} \int_{\Sigma} d^2 x \sqrt {\det g_{\mu\nu}} \tr F^2
}
=
\sum_{j} (2j + 1) \chi_{j} [U]
\e^{
- {g^2 {\cal A}} \, j(j + 1)/2
} \, ,
\end{equation}
where and $\chi_j [U]$ is its characters for the spin-$j$ representation
of the gauge group. The path integral representation is due to Ref.\
\cite{BlaTho91}, where the conjugation invariant delta function%
\footnote{It equates the eigenvalues of two unitary matrices.} is defined
by a group Fourier transform $\delta (U, U') = \sum_R \chi_R [U^{- 1}]
\chi_R [U']$. The disk transition amplitude (\ref{DiskAmplitude}) is used
to build the ones for arbitrary manifolds of genus $G$ by a gluing procedure
\cite{Rus90}. Eq.\ (\ref{DiskAmplitude}) reduces to the partition function
via (\ref{PartitionFunction}) by ${\cal Z} = \int d U {\cal Z} [U]$. Thus,
\be
{\cal Z} [U; g^2 {\cal A} = 2 \tau]
=
\vev{-v'| \e^{- \tau \mathcal{H}} |v}
\, ,
\label{Z-1}
\ee
c.f.\ Eq.\ (\ref{vev-1}), where $U = g_{-v'}^{-1} g_v^{\phantom{1}}$ and
$\tr \, [ U (\tilde\theta)] = 2 \cos\tilde\theta$ with $\,\tilde \theta =
\pi - \theta$ .
Thus, the one-loop cusp anomalous dimension \re{G0} is given by the integral
of the wave functional in two-dimensional Yang-Mills theory on the disc with
respect to its area
\be
{\mit\Gamma}_{\rm cusp} (\theta; \alpha_s)
=
-
\frac{\alpha_s N_c}{2 \pi}
\int_0^\infty d \tau
\Big(
{\cal Z} [U; 2 \tau] - {\cal Z} [\II; 2 \tau]
\Big)
\, .
\label{cusp=string}
\ee

As we have already mentioned, the transition amplitude of the particle on the
$SU(2)$ group manifold has two different representations, Eqs.~\re{theta} and
\re{vev-dual}. The first one coincides with \re{Z-1}. While the second one is
related to the  saturation of
the partition function on a Riemann surface by a sum over classical saddle
points in the path integral \cite{Wit92} (see also \cite{MinPol94}), ---  a
consequence of the Duistermaat-Heckman localization. It can be rephrased in physical
terms as instanton mechanism of confinement in two-dimensional Yang-Mills
theory \cite{GroMat94}. The instantons under consideration are solutions to
the Yang-Mills equations of motion on the two-dimensional disk with the
boundary conditions set by the holonomy $\tr \, U [A (x^0 = T, x^1)] = 2 \cos
\tilde\theta$. The classical configurations in the $A^0 (x^0, x^1)= 0$ gauge
correspond to straight paths connecting the initial and final points and read
in the topological charge-$\ell$ sector $A^1_\ell (x^0, x^1) = x^0 (\sigma_3/2)
(\tilde\theta + 2 \pi \ell)/{\cal A}$. The action evaluated on these instanton
solutions reads
\begin{eqnarray*}
S [A_\ell]
=
2 ( \tilde\theta +  2 \pi \ell )^2/(g^2 {\cal A})
\, ,
\end{eqnarray*}
and the weight factor $w_\ell = \tilde\theta + 2 \pi \ell$ in $w_\ell \exp\,
( - S_\ell )$. These properties are in a perfect agreement with our findings
in section \ref{PropagationSphere}, Eq.\ (\ref{vev-dual}).

It is well-known that the two-dimensional $SU (N)$ Yang-Mills theory is a string
theory \cite{Gro92,GroTay93}. Its partition function is given by the sum of maps
from two-dimensional worldsheet to two-dimensional target manifold $\Sigma$. The
explicit relation between the partition functions in two theories looks as
follows \cite{CorMooRam95}
\beq
\ln {\cal Z} [g^2 {\cal A}]
=
{\cal Z}_{\rm str}
\left[ g_s = 1/N, \alpha' = 1/(\pi g^2 N) \right]
\label{zz}
\eeq
with $N = 2$. Combining together Eqs.~\re{cusp=string} and \re{zz}, we
conclude that the one-loop cusp anomalous dimension \re{cusp=string} admits the
same stringy representation.

Eq.~\re{zz} is a counterpart of the relation between the transition amplitude
of firstly quantized particle and partition function of secondly quantized
field theory. Namely, it relates firstly quantized string with two-dimensional
Yang-Mills theory. According to \re{zz}, the partition function in the latter
theory is equal to the transition amplitude in the string theory with the
boundary conditions specified by the holonomy $U(\tilde\theta)$ on the disk
boundary. Moreover, one can establish a one-to-one correspondence between the
gauge invariant states in the Hilbert space of two-dimensional Yang-Mills
theory, $\tr \, [ U^n (\tilde\theta) ]$ with $n = 1, 2, \ldots$, and stringy
excitations (the oscillatory modes). The relation between the two looks as
follows \cite{Dou93,CorMooRam95}
\be
\tr \, [ U^n (\tilde\theta) ]
\to
\lr{ \alpha_n + \bar\alpha_{- n} } | \tilde\theta )
\, ,
\label{oscillator}
\ee
where the  $\alpha_n$ and
$\bar\alpha_n$ satisfy the commutation relations $[\alpha_n , \alpha_m] =
[\bar\alpha_n , \bar\alpha_m] = n\, \delta_{n + m,0}$ and $[\alpha_n ,
\bar\alpha_m] = 0$ and $| \tilde\theta )$ is the stringy coherent state
\begin{eqnarray*}
| \tilde\theta )
=
\exp
\left(
\sum_{n = 1}^\infty \e^{i n \tilde\theta} \frac{\alpha_{- n}}{n}
\right)
\exp
\left(
\sum_{n = 1}^\infty \e^{- i n \tilde\theta} \frac{\bar\alpha_{n}}{n}
\right)
| 0 \rangle
\, .
\end{eqnarray*}
The holonomy $U (\tilde\theta)$ is defined in the fundamental
representation of the $SU(2)$ group.%
\footnote{The closed string
picture for $SU(N_c)$ two-dimensional YM theory is valid
perturbatively only for $N_c\to\infty$. However, the stringy
representation exists for arbitrary $N_c$~\cite{BaeTay94}.}

We recall that $U (\tilde\theta) = g^{-1}_{-v'} g^{\phantom{1}}_v$
with $g_v$ given in \re{g_n} so that $\tr \, [ U^n (\tilde\theta)
] = 2 \cos (n \, \tilde\theta)$. Notice that $\tr \, [ U^n ]$ and
the characters $\chi_j [U]$ provide two linear independent bases
on the space of invariant functions on the $SU(2)$ group and,
therefore, they are related to each other by a linear
transformation \cite{BaeTay94}, a form of the Frobenius character
formula \cite{FulHar91}. This allows one to rewrite Eq.~\re{Z-1}
in terms of $\tr \, [ U^n ]$ and make use of \re{oscillator} in
order to re-express the transition amplitude for a particle on a
cylinder, deduced by gluing the opposite arcs of the boundary
holonomies of the disk amplitude \cite{Rus90},
\begin{equation}
{\cal Z} [U_1, U_2 ; 2 \tau]
=
\int d {\mit\Omega} \,
{\cal Z} [ {\mit\Omega} U_1 {\mit\Omega}^\dagger U_2 ; 2 \tau ]
\, ,
\end{equation}
in terms of a string amplitude
\be
{\cal Z} [U (\theta), U (0); 2 \tau]
=
\vev{-v'| \e^{- \tau \mathcal{H}} |v}
=
( \tilde\theta | \e^{-\tau \mathcal{H}_{\rm str}} | \tilde\theta = 0 )
\, ,
\label{Z=str}
\ee
where $| \tilde\theta = 0 )$ corresponds to a trivial holonomy $U(0) = \II$,
thus resulting merely to a disk amplitude ${\cal Z} [U (\tilde\theta), U (0);
2 \tau] = {\cal Z} [U (\tilde\theta); 2 \tau]$. The group theory Hamiltonian
$C_2 (R)$ is not diagonal in the basis spanned by string states and involves
splitting and joining of strings. The string Hamiltonian is defined for the
$SU(2)$ group as \cite{MinPol93}
\beq
\mathcal{H}_{\rm str}
=
2 ( L_0 + \bar L_0 )
+
\frac{1}{2}( L_0 - \bar L_0 )^2
+
( V + \bar V )
\, ,
\eeq
where the Virasoro generator and interaction vertex operator read
\begin{eqnarray*}
L_0 = \frac{1}{2} \sum_n \alpha_{- n} \alpha_n
\, , \qquad
V = \frac{1}{2} \sum_{n, m > 0}
( \alpha_{- n - m} \alpha_n \alpha_m + \alpha_{- n} \alpha_{- m} \alpha_{n + m} )
\, ,
\end{eqnarray*}
with $\bar L_0$ and $\bar V$ given by similar expressions. Substituting
\re{Z=str} into \re{cusp=string} one could obtain the representation for one-loop
cusp anomalous dimension in terms of string propagator
\be
{\mit\Gamma}_{\rm cusp} (\theta; \alpha_s)
=
- \frac{\alpha_s N_c}{2\pi}
\left[
( \tilde\theta | 1/{\mathcal{H}_{\rm str}} | 0 )
-
( 0 | 1/{\mathcal{H}_{\rm str}} | 0 )
\right] \, .
\ee

We would like to stress that the string corresponding to the one-loop cusp anomaly
is not of the Nambu-Goto type \cite{CorMooRam95}. At the same time, the cusp anomaly
at strong coupling is described by the Nambu-Goto string on the AdS background whose
tension scales as $\lr{\alpha_s N_c}^{1/2}$ (see next section). Going over from weak
to strong coupling regime one expects to find the transition from the former string
to the latter. The mechanism governing such transition remains unclear. One of
possible scenarios was proposed in Ref.\ \cite{horava}. It is based on the
identification of the two-dimensional Yang-Mills theory \re{Z-2D}, rewritten as
\begin{eqnarray*}
- \frac{1}{g^2} \tr F^2
\to
\tr \, [ 2 F \phi + g^2 \phi^2 ]
\, ,
\end{eqnarray*}
as a topological string (for $g^2 = 0$) perturbed by a rigidity
term \cite{Pol-book,Pol86}. The Nambu-Goto action was conjectured
to arise through the dimensional transmutation mechanism at strong
coupling.

It is worth mentioning on the  relation
of Yang-Mills theory with AdS background. Its appearance can be
understood by noticing that after analytical continuation of the
cusp anomaly from the Euclidean to Minkowski kinematics, one has
to deal with a two-dimensional Yang-Mills theory with the
$SL(2,\mathbb{R})$ gauge group. It is well-known \cite{FukKam85}
that at $g^2 = 0$ such theory is equivalent to topological
Jackiw-Teitelboim gravity with the action
\begin{equation}
S
=
\int d^2x \sqrt{\det g_{\mu\nu}} \, \left( R(g) - \Lambda \right) \eta
\, ,
\end{equation}
where $\eta$ is the dilaton field and $\Lambda$ is the cosmological constant.
Solutions to the classical equations of motion give rise to the AdS$_2$
gravity coupled to the dilaton.

\section{Cusp anomaly at strong coupling}
\label{StrongCoupling}

The main goal of our previous discussion of the cusp anomaly at
weak coupling was to emphasize its quantum nature as a transition
amplitude for a particle on the AdS${}_3$ space. In this section,
we will argue that at strong coupling the cusp anomaly is given by
Hamilton-Jacobi action function corresponding to a classical
mechanical system defined on the same space.

According to the AdS/CFT correspondence \cite{Mal97,GunKlePol98,Wit98}, the
strong coupling regime in gauge theories is related to the supergravity
limit of a string theory on the AdS${}_5\times$S${}^5$ background. In the
present discussion we are interested in operators with large angular momentum
$J$ where the conventional (supergravity field)/(Yang-Mills operator)
correspondence is not applicable, and one has to solve the string theory
semiclassically \cite{Mal98,GubKlePol02}. For the light-cone observables
discussed here, like quark distribution functions \re{CollinearPDFs} and
light-like Wilson loops \re{LargeX}, the full conformal QCD group $SO(4,2)$
is effectively reduced to its collinear subgroup $SU(1,1)$. It is only the
latter which acts non-trivially on the field operators ``living'' on the
light-cone. The group $SL(2, \mathbb{R}) \times SL(2, \mathbb{R})$ is an
isometry of the AdS$_3$. Therefore, applying the gauge/string correspondence,
instead of the full AdS$_5$ space it will be enough to consider only its
AdS$_3$ subspace.

Let us remind a few elementary facts about anti-de Sitter space. The AdS$_3$
space with the Lorentzian signature is a hypersurface embeded in flat
$\mathbb{R}^{2,2}$
\begin{equation}
\label{AdSembedding}
X_0^2 - X_1^2 - X_2^2 + X_3^2 = R^2 \, .
\end{equation}
We set its radius to be $R = 1$ for simplicity. The $SU (1,1)$ group structure
becomes manifest via the following parametrization
\begin{equation}
g =
\left(\!\!
\begin{array}{c}
X_0 + i X_3 \ X_1 - i X_2 \\
X_1 + i X_2 \ X_0 - i X_3
\end{array}
\!\!\right)
= \sum_{a = 0}^3 X_a \tau_a
\, ,
\label{surface}
\end{equation}
with $\tau_a = \{ \II , \sigma_1 , \sigma_2 , i \sigma_3 \}$. To parametrize the
hypersurface \re{AdSembedding}, we will use two different sets of coordinates
\cite{AhaGubMalOogOz99}. In the global $(\rho,\tau,\phi)$-coordinates the AdS
space looks like
\begin{equation}
X_0 + i X_3 = \cosh \rho \ {\rm e}^{i \tau}
\, , \qquad
X_1 + i X_2 = \sinh \rho \ {\rm e}^{i \phi}
\, .
\end{equation}
and in local Poincar\'e $(u,x,t)$-coordinates
\be
X_0 + X_1 = u
\, , \qquad
X_2 = u x
\, , \qquad
X_3 = u t
\, .
\ee
The transformation from one set of coordinates to the other is achieved via
the map
\begin{eqnarray}
u \!\!\!&=&\!\!\!
\cosh\rho \ \cos \tau + \sinh\rho \ \cos \phi
\, , \nonumber\\[2mm]
x \!\!\!&=&\!\!\!
\frac{\sinh \rho \ \sin\phi}{\cosh\rho \ \cos \tau + \sinh\rho \ \cos \phi}
\, , \nonumber\\
t \!\!\!&=&\!\!\!
\frac{\cosh \rho \ \sin\tau}{\cosh\rho \ \cos \tau + \sinh\rho \ \cos \phi}
\, .
\label{sets}
\end{eqnarray}
For the discussion which follows we introduce the Rindler coordinates on
the conformally flat part $\mathbb{R}^{1,1}$ of AdS$_3$ in Poincar\'e
parametrization
\begin{equation}
\label{Rindler}
t = r \cosh \theta \, , \qquad x = r \sinh \theta \, .
\end{equation}
The metric on the AdS space in the global coordinates reads, see e.g.,
\cite{AhaGubMalOogOz99},
\be
\label{ds1}
ds^2
=
\frac{1}{2} \tr \left( g^{-1} d g \right)^2
=
- \cosh^2\!\rho \ d \tau^2 + \sinh^2\!\rho \ d\phi^2 + d \rho^2
\ee
and in the Poincar\'e-Rindler coordinates
\be
\label{ds2}
ds^2
=
\frac{du^2}{u^2} + u^2 \left( dx^2 - dt^2 \right)
=
\frac{du^2}{u^2} + u^2 \left(-d r^2 + r^2 d \theta^2 \right)
\, .
\ee
Here, the two sets of coordinates have different physical interpretation. In
Eq.~\re{ds1}, $0\le\rho<\infty$ defines the radial coordinate on the AdS space,
$0\le \phi < 2 \pi$ is the azimuthal angle and $- \infty < \tau < \infty$ sets
up the AdS time. Notice that the latter is different from the time variable in
\re{ds2}. In Eq.~\re{ds2}, $0 \le u^2 < \infty$ is the Liouville coordinate, $t$
and $x$ are the time and the spacial coordinates, respectively, on the
hyperplane in Minkowski space to which the contour entering the definition of
the Wilson loop \re{W} (see also Fig.~\ref{Cylinder}) belongs to. In the polar
coordinates, choosing $r = 0$ at the cusp, one identifies $\theta$ in \re{ds2}
as the cusp angle. Since the relation \re{sets} between two sets of the
coordinates is nonlinear, the same trajectory of a test particle on the AdS
space looks differently in the $(\rho,\phi,\tau)$ and $(r,\theta,u)$ coordinates.

Let us now turn to the analysis of the cusp anomaly in the strong coupling
regime. As was mentioned in section \ref{QCDptTheory}, there are two apparently
different approaches to calculate the cusp anomalous dimension within the gauge/string
correspondence. One of them relies on the relation between the large-$J$
behavior of the anomalous dimensions of twist-two composite operators and cusp
anomaly, Eq.~\re{gamma-as}. In this way, following \cite{GubKlePol02}, one can
calculate  ${\mit\Gamma}_{\rm cusp}(\alpha_s)$ as the energy minus spin of a
folded closed string rapidly rotating in the AdS space in the $(\rho,\phi,\tau)$
coordinates, Eq.~\re{GKP}. In the second approach \cite{DruGroOog99,Kru02,Mak02},
one calculates ${\mit\Gamma}_{\rm cusp}(\alpha_s)$ from the minimal surface of
the worldsheet of an open string propagating in the AdS space in the $(r,\theta,u)$
coordinates with the ends sliding along two rays at the boundary $u = \infty$
with the cusp angle $\theta\to\infty$. In both approaches, the result for the
cusp anomaly is expressed in terms of solutions to classical equations of
motion. The main difference between the two cases is the form of the classical
Hamiltonian and underlying picture of classical motion.%
\footnote{Another difference is that the calculation of Ref.\ \cite{GubKlePol02}
can be performed only in the AdS$_3$ space with Lorentzian signature, whereas in
the case of the Wilson loop, Refs.\ \cite{DruGroOog99,Kru02,Mak02}, the result
can be reproduced by an analytical continuation from Euclidean space.} In this
section, we will demonstrate the equivalence between the two approaches.

\subsection{Open string and Wilson loop}
\label{OpenSection}

\begin{figure}[t]
\unitlength1mm
\begin{center}
\mbox{
\begin{picture}(0,32)(85,0)
\put(10,0){\insertfig{15}{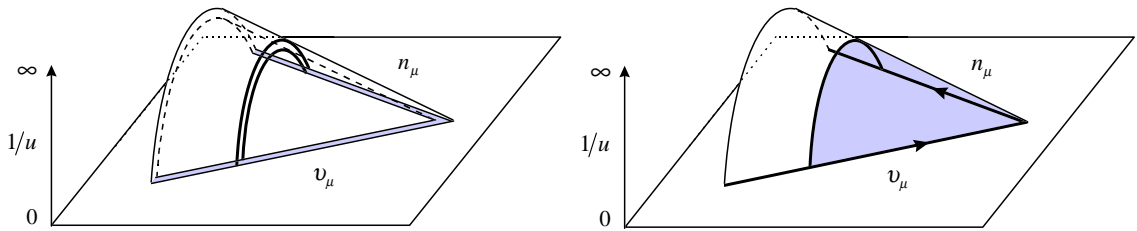}}
\end{picture}
}
\end{center}
\caption{\label{MinimalSurface} The minimal surfaces swept by open strings
whose ends move along the contour of the Wilson loop. The surface of the
Wilson loop in the adjoint representation (left) is a double of the Wilson
loop in the fundamental representation (right).}
\end{figure}

To begin with, we recall the calculation of a Wilson loop with a cusp. The
Nambu-Goto action for a string propagating in the AdS${}_3$ target space looks
like
\be
S = 2 \, \frac{R^2}{2\pi\alpha'} \,
\int d^2 \sigma \sqrt{ -\det\| G_{MN} \partial_a X^M \partial_b X^N \|}
\, ,
\label{Nambu-Goto}
\ee
where $R^2/\alpha'=\sqrt{g^2N_c}$ and $G_{MN}$ is the metric tensor on the
$AdS_3$ space, $ds^2 = G_{MN} d X^M d X^N$, Eqs.~\re{ds1} and \re{ds2}. The
Wilson loop is defined by a classical configuration that minimizes this action,
$W \sim \exp(i S_{\rm min})$. The additional factor $2$ in the right-hand side
of \re{Nambu-Goto} takes into account that the Wilson loop is taken in the
adjoint representation of the $SU(N_c)$ and, in multi-color limit $N_c \to \infty$,
it is just the square of the same loop in the fundamental representation. Also,
the minus sign under the square-root in \re{Nambu-Goto} ensures that $S$ is real
in Minkowski signature.

The main contribution to the cusp anomaly comes from the vicinity of the cusp,
see Fig.\ \ref{MinimalSurface}. In the Poincar\'e-Rindler $(r,\theta,u)$
coordinates the minimal surface can be written as \cite{DruGroOog99}
\be
u = \frac{f(\theta)}{r} \, .
\label{u}
\ee
Choosing $(\sigma_1, \sigma_2) = (\theta,r)$ as local coordinates on the
worldsheet, one finds the induced metric as
\be
G_{MN} \partial_a X^M \partial_b X^N
=
\left(
\begin{array}{cc}
({\dot f}/f)^2 + f^2 & - {\dot f}/(f r) \\
- {\dot f}/(f r)     & (1 - f^2)/r^2
\end{array}
\right)
\, ,
\ee
where ${\dot f}=\partial f(\theta)/\partial\theta$. Then, the
action becomes
\be
S_{\rm min} = 2\sqrt{\frac{\alpha_s N_c}{\pi}}\int\frac{d r}{r} \int d
\theta \sqrt{{\dot f}^2 - f^2 + f^4} \equiv i{\mit\Gamma}_{\rm
cusp}(\theta,\alpha_s)\ln\frac{r_{\rm max }}{r_{\rm min }} \, ,
\label{S-cusp}
\ee
where the cusp anomalous dimension is given by
\be
{\mit\Gamma}_{\rm cusp}(\theta; \alpha_s)
=
2\sqrt{\frac{\alpha_s N_c}{\pi}}
\int_0^\theta d \theta \, \sqrt{-{\dot f}^2 + f^2 - f^4}
\, .
\ee
This allows us to interpret ${\mit\Gamma}_{\rm
cusp}(\theta,\alpha_s)$ as a classical action of a particle with
the Lagrangian $\mathcal{L} [f]=(-{\dot f}^2 + f^2 - f^4)^{1/2}$,
where $f(\theta)$ and $\theta$ play the r\^ole of the coordinate
and the time, respectively. The energy $E$ and momentum
$P(\theta)$ of the particle take the form
\be
E = {\dot f} \frac{\partial \mathcal{L}[f]}{\partial {\dot f}} - \mathcal{L} =
\frac{-f^2 + f^4}{\sqrt{-{\dot f}^2 + f^2 - f^4}}
\, , \qquad
P
=
\frac{\partial \mathcal{L}[f]}{\partial {\dot f}}
= -\frac{{\dot f}}{\sqrt{ -{\dot f}^2 + f^2 - f^4}}\,.
\ee
Being the integral of motion, $E = \rm const$, the energy determines the solutions
to the classical equations of motion. The action calculated along classical
trajectory satisfies the Hamilton-Jacobi equation and is given by
\be
{\mit\Gamma}_{\rm cusp}(\theta; \alpha_s)
=
2 \sqrt{\frac{\alpha_s N_c}{\pi}}
\int_0^\theta d \theta \, \lr{P(\theta) {\dot f}(\theta) - E}
\, .
\label{cusp-str}
\ee
Since the minimal surface ends at the boundary of the AdS space, $u = \infty$,
the classical trajectories have to satisfy the boundary condition $f(0) =
\infty$. As was shown in Refs.\ \cite{Kru02,Mak02}, the asymptotic behavior of
the cusp anomaly at large $\theta$ is governed by the contribution of classical
solutions with the energy $E = -1/2$. In this
case, the classical trajectory
starts at infinity, $f(0) = \infty$, and approaches $f(\theta) \to 1/\sqrt{2}$ as
$\theta \to \infty$. Since $P(\theta) \sim \dot f(\theta)$ vanishes in this
limit, $\dot f = 0$, one finds from \re{cusp-str}
\be
{\mit\Gamma}_{\rm cusp}(\theta; \alpha_s) = \sqrt{\frac{\alpha_s N_c}{\pi}}
\, \theta
\, ,
\label{Gamma-str}
\ee
in agreement with \re{light-cusp} and \re{cusp-strong}. The corresponding minimal
surface can be translated into the global coordinates by noticing that $f^2 (\theta)
= X_3^2 - X_2^2\to 1/2$. Then, one gets from the equation for
AdS embedding (\ref{AdSembedding}),
\begin{equation}
\label{Surface}
X_3^2 - X_2^2 = 1/2
\, , \qquad
X_0^2 - X_1^2 = 1/2
\, ,
\end{equation}
which is the result of Ref.\ \cite{Kru02}.

\subsection{Rotating closed string}
\label{RotationClosed}

Let us now recapitulate the consideration of a rotating folded closed string in
the AdS${}_3$ space around its center-of-mass following \cite{GubKlePol02}, see
also \cite{FolLowGkp02}. In distinction with the previous case, one chooses to
work in the global $(\rho,\phi,\tau)$-coordinates (\ref{ds1}) and assumes that
the center of the string lies at $\rho=0$. The string action is given by the
same Nambu-Goto expression \re{Nambu-Goto} but with different boundary
conditions. Choosing the gauge $\sigma_1 = \tau$ and $\sigma_2 = \rho$, one
finds the induced metric as
\be
G_{MN} \partial_a X^M \partial_b X^N
=
\left(
\begin{array}{cc}
-\cosh^2\rho + {\dot \phi}^2 \, \sinh^2 \rho & 0 \\
0 & 1
\end{array}
\right) \, ,
\ee
where $\phi = \phi(\tau)$ is an azimuthal angle of a point on the string
with the AdS time and radial coordinates, $\tau$ and $\rho$, respectively,
and ${\dot\phi}\equiv\partial\phi/\partial\tau$ the corresponding angular
velocity. As a consequence, the action of the rotating stretched string looks
like
\be
S
=
4 \, \frac{R^2}{2\pi\alpha'} \,
\int d \tau \int_0^{\rho_0} d\rho \,
\sqrt{\cosh^2\rho  - {\dot \phi}^2(\tau) \sinh^2 \rho}
\equiv
\int d \tau \, \mathcal{L} [\phi]
\, .
\label{S-GKP}
\ee
Here the additional factor $4$ counts the number of segments of the folded string
rotating around $\rho = 0$ and the maximal radial coordinate $\rho \le \rho_0$ is
determined by
\be
\coth^2\rho - {\dot \phi}^2(\tau) \ge 0
\, .
\ee
Eq.~\re{S-GKP} defines a classical mechanical model of a rotating rod with the
Lagrangian $\mathcal{L} [\phi]$. Its energy and angular momentum are
\ba
\label{E-P-rho0}
E
\!\!\!&=&\!\!\!
{\dot \phi} \frac{\partial \mathcal{L} [\phi]}{\partial {\dot\phi}}
-
\mathcal{L}
=
- 4 \sqrt{\frac{\alpha_s N_c}{\pi}}
\int_0^{\rho_0} d \rho
\frac{\cosh^2 \rho}{\sqrt{\cosh^2 \rho - {\dot \phi}^2 \sinh^2 \rho}}
\, , \\
J
\!\!\!&=&\!\!\!
\frac{\partial \mathcal{L}[\phi]}{\partial {\dot \phi}}
= - 4 \sqrt{\frac{\alpha_s N_c}{\pi}}
\int_0^{\rho_0} d \rho
\frac{{\dot \phi}\, \sinh^2 \rho}{\sqrt{\cosh^2 \rho - {\dot \phi}^2 \sinh^2\rho}}
\, .
\ea
Both quantities are integrals of motion so that the classical trajectories
are specified by the values of $E$ and $J$. The action \re{S-GKP} evaluated
along the classical trajectory is given by\footnote{Note that the pair
$(J,\phi)$ defines the action-angle variables for the system under
consideration.}
\be
S_{\rm cl} =
\int d \tau \left( J \, {\dot \phi} - E \right)
=
2{\gamma}_{J} (\alpha_s)\ln \frac{r_{\rm max}}{r_{\rm min}}
\, ,
\label{gamma-2}
\ee
where  $\tau_{\rm max/\min}=\ln r_{\rm max/\min}$ and
\be
{\gamma}_{J}(\alpha_s)
=
\frac{1}{2} \int_0^{2\pi} \frac{d \tau}{2 \pi}
\left( J \, {\dot \phi} - E \right)
=
\frac12 \lr{- E + J\, \omega}
\label{E-J}
\ee
with $\omega = \dot\phi$ being the angular velocity of the rod. The additional
factor $2$ in the right-hand side of \re{gamma-2} counts the number of end-points
of the folded string. The anomalous dimension defined in this way is the
coefficient in front of the AdS time in the expression for the action function.
The latter is the solution to the Hamilton-Jacobi equations for the system
\re{S-GKP}. In the limit of long strings,
\be
\rho_0 = 1/2 \ln(1/\eta)\gg 1
\, , \qquad
\omega = 1 + 2 \eta
\, ,
\label{rho0}
\ee
for $\eta \to 0$, one finds the energy and angular momentum of the folded
string as
\be
E
=
2 \sqrt{\frac{\alpha_s N_c}{\pi}}\lr{\eta^{-1} - \ln\eta}
\, , \qquad
J
=
2 \sqrt{\frac{\alpha_s N_c}{\pi}}\lr{\eta^{-1} + \ln\eta}
\, .
\ee
Substituting these relations into \re{E-J}, one obtains
\be
\label{StrongCouplingTwo}
\gamma_J(\alpha_s)
=
2 \sqrt{\frac{\alpha_s N_c}{\pi}}\ln J
\, .
\ee
This expression defines the anomalous dimension of the twist-two operators
$\widehat O_J(0)$ at strong coupling.

\subsection{Multi-particle operators: minimal surfaces}

As we have seen in the previous sections, the anomalous dimensions of the twist-two
operators at strong coupling can be obtained using two different approaches
based on the calculation of the Wilson loop with a cusp and the classical
energy of a long string rotating on the AdS background. In this section we will
generalize these results to $N$-particle conformal operators of higher twist.

We have demonstrated in section \ref{WilsonApp}, that the anomalous dimensions of
such operators at large spins $J$ occupy the band \re{bounds} whose boundaries,
Eqs.~\re{gamma-max} and \re{gamma-min}, are defined by the cusp anomalous
dimension. Since this result holds for arbitrary coupling constant, one makes use
of \re{cusp-strong} to replace ${\mit\Gamma}_{\rm cusp}(\alpha_s)$ by its asymptotic
behavior at strong coupling. Remarkably enough, the same result can be obtained
from the gauge/string duality.

Following the approach described in section \ref{OpenSection}, one has to
construct the minimal surface on the AdS${}_3$ target space whose boundary
involves multiple cusps. The number of cusps, $k$, varies along the band. On the
upper and lower boundary it equals $N$ and $2$, respectively. At large $N_c$,
the expectation value of the product of Wilson loops factorizes into the product
of their expectation values (see Fig.\ \ref{CylinderTopology})
\be
\vev{\tr W_\Pi[\xi_1,\xi_2]\ldots \tr W_\Pi[\xi_k,\xi_1]}=
\vev{\tr W_\Pi[\xi_1,\xi_2]}\ldots \vev{\tr W_\Pi[\xi_k,\xi_1]}
\label{W-vev}
\ee
This implies that the area of the minimal surface corresponding to the Wilson
loop with $k$ cusps is given by the sum of $k$ elementary areas derived in
section \ref{OpenSection}
\be
\vev{\tr W_\Pi[\xi_j,\xi_{j+1}]}
\sim
\exp\left( S(\theta_j) + S (\theta_{j+1}) \right)
=
\mu^{
{\mit\Gamma}_{\rm cusp} (\theta_j;\alpha_s)
+
{\mit\Gamma}_{\rm cusp} (\theta_{j+1}; \alpha_s)
}
=
\mu^{
(\theta_j + \theta_{j+1}) {\mit\Gamma}_{\rm cusp} (\alpha_s)
}
\, .
\ee
Substituting this relation into \re{W-vev}, we calculate the total area of the
minimal surface
\be
S (\theta_1, \ldots, \theta_k)
=
2 \sum_{j=1}^k S (\theta_j)
=
2 {\mit\Gamma}_{\rm cusp} (\alpha_s)
\sum_{j=1}^k \theta_j \ln\mu
\sim
{}2 k \theta {\mit\Gamma}_{\rm cusp} (\alpha_s) \ln\mu
\ee
for $\theta_1\sim \ldots \sim \theta_k \sim \theta \gg 1$. As before, the
coefficient in front of the $\ln\mu$ at $k = 2$ and $k=N$ can be identified as
the anomalous dimensions of the $N$-particle conformal operators,
$\gamma_J^{\rm min}$ and $\gamma_J^{\rm max}$, respectively, for $\theta\sim J$.

\subsection{Multi-particle operators: revolving closed string}

Let us turn to the picture of a rotating closed string. We remind that the
anomalous dimension of the twist-two conformal operators, $\gamma_J^{\rm tw-2}
(\alpha_s)$, at strong coupling is related to the energy of the rotating
Nambu-Goto string evaluated on a classical configuration with the minimal energy
for a given angular momentum $J\gg 1$. In the AdS background such configuration
corresponds to a folded rotating long string. It is worth mentioning that the
emerging picture is a generalization of the well-known hadronic string for the
meson states from flat to curved background, see, e.g., \cite{Art75}. Attempting
to extend the Gubser-Klebanov-Polyakov approach to $N$-particle conformal
operators, one immediately encounters the following difficulty. In distinction
with the $N=2$ case, the anomalous dimensions occupy the band \re{bounds}. On the
stringy side, this indicates the existence of additional stringy degrees of
freedom. One expects that their total number should be $N-2$ in accordance with
the total number of integrals of motion $q_2, \ldots,q_N$ in the Heisenberg
spin chain (both in classical and quantum cases). The spectrum of the integrals
of motion is specified by the total angular momentum $J$ and the set of integers
$\ell_1,\ldots,\ell_{N-2}$.%
\footnote{These integers appear in the Bohr-Sommerfeld quantization conditions
imposed on the orbits of classical motion.}
Going over to the strong coupling regime, the integrability properties of the
evolution equations may be lost, but the analytical structure of the energy
spectrum remains intact. In other words, for large $J$ the anomalous dimensions
of $N$-particle operators are parameterized by the same set of integers,
$\gamma_J = \gamma_J (\alpha_s; \ell_1, \ldots, \ell_{N-2})$ although the
explicit form of this dependence may be different at strong and weak coupling.
As we will argue below, these additional degrees of freedom can be identified
as {\sl string junctions\/}.

\begin{figure}[t]
\unitlength1.5mm
\begin{center}
\mbox{
\begin{picture}(0,27)(70,0)
\put(30,0){\insertfig{12}{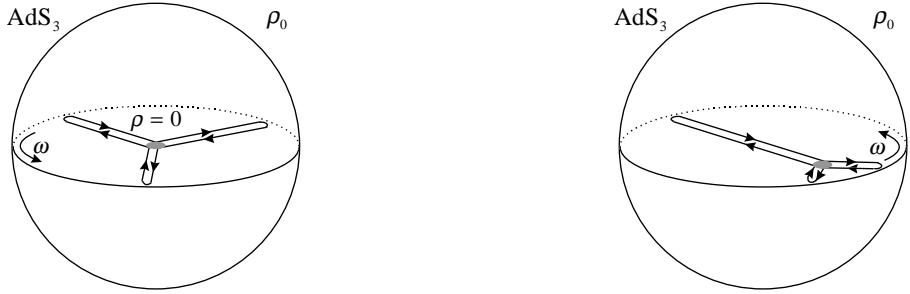}}
\end{picture}
}
\end{center}
\caption{\label{Baryon} Baryon string with a junction.}
\end{figure}

Let us consider the simplest case of the $N=3$ conformal operators. Similar to
the $N=2$ case, we expect to recover a folded closed string rotating on the AdS
background. Its total angular momentum equals the Lorentz spin of the conformal
operator. Since the logarithmically enhanced contribution, $\sim \ln J$, to the
energy of the string originates from the boundary region \re{rho0}, the large $J$
asymptotics of the anomalous dimension depends on how many bits of the folded
string approach the boundary (see Fig.\ \ref{Baryon}). The fact that the
anomalous dimensions on the upper boundary \re{gamma-max} scale as $3\ln J$
suggests that the corresponding stringy configuration consists of three long
bits, which are interconnected at some point close to the center of the AdS.
Similar {\sf Y}-shaped configurations are well-known in QCD as describing
baryonic string and following \cite{Art75} we will refer to the string vertex as
the string junction. There is however a number of important differences.

Since the quarks in QCD belong to the fundamental representation of the $SU(N_c)$
group for $N_c=3$, the color-singlet baryonic operators are built from $N_c$
quarks and the baryonic vertex in the corresponding {\sf Y}-shaped hadronic
string contains the same number of string bits. Within the AdS/CFT framework, at
large $N_c$, similar baryonic vertices have been constructed in Ref.~\cite{Wit98b}.
In supersymmetric theories, fermions belong to the adjoint representation of the
$SU(N_c)$, which allows one to construct color-singlet composite operators
containing an arbitrary number $N \ge 2$ of fermions, see e.g. Eq.~\re{O-N}. This
leads to important differences in the renormalization properties of such
operators both at the weak and strong coupling. Namely, at large $N_c$, the
interaction between fermions in the adjoint representation occurs only between
nearest neighbours while in the fundamental representation, due to antisymmetry
of the baryonic vertex under permutation of quarks, the interaction between any
pair of quarks is allowed. In the string representation, one can effectively
replace a Wilson line in the adjoint representation by a pair of Wilson lines in
the fundamental representation running in opposite directions. Then, one can
construct the {\sf Y}-shaped baryonic vertex as shown in Fig.\ \ref{Baryon}. In
distinction with the previous case, this vertex contains six string bits, but,
as before, we shall call it the string junction.

The {\sf Y}-shaped string on the AdS background is described by the
action
\be
S = 2 \, \frac{R^2}{2\pi\alpha'} \,
\sum_{k = 1}^{N = 3}
\int d \sigma_1^{\scriptscriptstyle (k)} d \sigma_2^{\scriptscriptstyle (k)}
\sqrt{ - \det \|
G^{MN}
\partial_{a} X_M^{\scriptscriptstyle (k)}
\partial_{b} X_N^{\scriptscriptstyle (k)}
\|}
\, ,
\label{ThreeString}
\ee
where the superscript $(k)$ enumerates three different ``arms'' and
$\sigma_a^{\scriptscriptstyle (k)}$ are local coordinates on the worldsheet of
the $k$-th arm. Making use of the reparameterization invariance and assigning
$\sigma_2^{\scriptscriptstyle (k)} = 0$ and $\sigma_2^{\scriptscriptstyle (k)} =
\pi$ to the folding point and string junction, respectively, one can write the
boundary condition along the string junction as $ X_M^{\scriptscriptstyle (1)}
(\sigma_1^{\scriptscriptstyle (1)}, \pi) = X_M^{\scriptscriptstyle (2)}
(\sigma_1^{\scriptscriptstyle (2)}, \pi) = X_M^{\scriptscriptstyle (3)}
(\sigma_1^{\scriptscriptstyle (3)}, \pi) $. The string equations of motion
corresponding to the action \re{ThreeString} take the following form in the
conformal gauge \cite{Art75,deVega,Kri94}
\be
\partial_+ \partial_- X_M^{\scriptscriptstyle (k)}
+
\left(
\partial_+ X^{\scriptscriptstyle (k)} \cdot \partial_-
X^{\scriptscriptstyle (k)} X_M^{\scriptscriptstyle (k)}
\right)
= 0
\, , \qquad
T_{\pm \pm}^{\scriptscriptstyle (k)}
=
\left(
\partial_\pm X^{\scriptscriptstyle (k)} \cdot \partial_\pm X^{\scriptscriptstyle (k)}
\right)
= 0
\, ,
\label{EQs}
\ee
where $(X \cdot Y) \equiv G^{MN} X_M Y_N$ is the scalar product on the AdS space
and $\partial_\pm \equiv \partial/\partial \sigma_\pm$ with $\sigma_\pm = \sigma_1
\pm \sigma_2$ being light-cone coordinates on the world-sheet. Eqs.~\re{EQs} are
invariant under reparametrization $\sigma_\pm \to f (\sigma_\pm)$. To fix this
ambiguity we identify the local coordinate on the world-sheet as the time
coordinate on the AdS space $\sigma_1 = \tau$. Then, the energy and the angular
momentum of the string are given by
\ba
E\!\!\!&=&\!\!\!
2 \, \frac{R^2}{2\pi\alpha'} \,
\sum_{k = 1}^3
\int_0^{\beta_k (\tau)} d \sigma_2^{\scriptscriptstyle (k)}
G_{\tau\tau} (X^{\scriptscriptstyle (k)})
\nonumber\\
J\!\!\!&=&\!\!\!
2 \, \frac{R^2}{2\pi\alpha'} \,
\sum_{k = 1}^3
\int_0^{\beta_k(\tau)} d \sigma_2^{\scriptscriptstyle (k)}
\dot\phi_k G_{\phi\phi}(X^{\scriptscriptstyle (k)})
\label{E,J}
\ea
where $G_{\tau\tau}=\cosh^2 \rho_{\scriptscriptstyle (k)} (\tau)$ and
$G_{\phi\phi} = \sinh^2 \rho_{\scriptscriptstyle (k)} (\tau)$ define the
AdS${}_3$ metric \re{ds1}, while $\rho_{\scriptscriptstyle (k)}$ and
$\phi_{\scriptscriptstyle (k)}$ are the radial and angular AdS coordinates of the
$k$-th arm. Notice that in this gauge the local parameters $\sigma_2$
corresponding to each arm take the same value at the folding point,
$\sigma_2^{\scriptscriptstyle (k)} = 0$ and different values at the junction
$\sigma_2^{\scriptscriptstyle (k)}=\beta_k (\tau)$. To find the explicit form of
$\beta_k(\tau)$ one has to impose the junction conditions. In covariant form they
take the form \cite{Art75}
\be
\sum_{k = 1}^3 \frac{d X_M^{\scriptscriptstyle (k)} (\tau, \beta_k (\tau))}{d\tau}
=
\sum_{k = 1}^3
\lr{
\partial_\tau  + \dot\beta_k (\tau)\partial_{\sigma_2}
}
X_M^{\scriptscriptstyle (k)} (\tau,\sigma_2) \bigg|_{\sigma_2 = \beta_k (\tau)}
= 0
\, ,
\label{EQ-junction}
\ee
together with
\be
X_M^{\rm (junction)} (\tau)
\equiv
X_M^{\scriptscriptstyle (1)} (\tau,\beta_1(\tau))
=
X_M^{\scriptscriptstyle (2)} (\tau,\beta_2(\tau))
=
X_M^{\scriptscriptstyle (3)} (\tau, \beta_3(\tau))
\, .
\label{wl-junction}
\ee
Solving the system of equations \re{EQs}, \re{EQ-junction} and \re{wl-junction},
one can find the classical motion of the {\sf Y}-shaped string on the AdS
background and apply \re{E,J} to calculate the corresponding energy and angular
momentum.

In the flat space, for hadronic QCD string, this analysis has been carried out in
Refs.\ \cite{Art75}. In that model, the {\sf Y}-shaped string describes the
spectrum of baryons and the string junction plays the r\^ole of their additional
degree of freedoms. It was found that the dependence of the mass of baryons, $m =
E$, on their angular momentum $J$ takes the Regge form, $J/m^2 = \alpha'/\kappa$
with $\kappa$ depending on the classical dynamics of the junction and taking the
values within the band $2\le \kappa \le 3$. The maximal value $\kappa = 3$
corresponds to the configuration when the string junction is at rest and three
bits of the string have the form of the rods of the same length, rotating with
the same angular velocity and forming the same angle $2 \pi/3$ against each other
(recall the analogy with the interior minimal surface of three joint soap
bubbles). The minimal value $\kappa = 2$, corresponds to the meson-like Regge
trajectory, i.e.\ $N = 2$ in Eq.\ (\ref{ThreeString}). In this case, the baryon
has the diquark-quark structure, that is the end-points of two bits are located
close to each other and to the string junction. As we will argue in a moment,
similar picture emerges in the AdS geometry.

To start with, we notice that short strings rotating around the center of the
AdS${}_5$ do not feel its curvature and, therefore, look the same as strings in
a flat background. That is, the dependence of the energy, $E$, and the angular
momentum, $J$, of the string on its angular velocity $\omega$ is the same in
two cases. The only difference is due to different representation of the fermions
-- the open hadronic string in QCD is replaced by a folded closed string in the
supersymmetric case. In the long-string limit, as was first shown in Ref.\
\cite{GubKlePol02}, a long folded string rotating on the AdS background gives
rise to the anomalous dimensions of local composite operators of large spin
$J$. According to \re{GKP}, the anomalous dimension scales as $\ln J$ with the
prefactor depending on the number of string bits reaching the boundary \re{rho0}.
In particular, meson-like long folded string gives rises rise to the anomalous
dimension of two-particle conformal operators, Eq.\ (\ref{StrongCouplingTwo}).

Generalizing this picture to the case of baryon-like folded strings, we consider
the same {\sf Y}-shaped configuration as in the hadronic string with the only
difference that each bit of the ``fundamental'' string is replaced by two bits
of the folded string. One can verify that such configuration satisfies the
classical equations of motion on the AdS background. Since the string junction
vertex is at rest, the energy of the rotating folded {\sf Y}-shaped string is
given by the sum of energies of three arms, i.e.\ we can choose the gauge
$\sigma_{1\scriptscriptstyle (k)} = \tau_{\scriptscriptstyle (k)}$ and
$\sigma_{2\scriptscriptstyle (k)} = \rho_{\scriptscriptstyle (k)}$. The same is
true for the total angular momentum. Due to symmetry of the configuration, three
arms have the same energy and the angular momentum which in their turn are equal
to half of the energy and the angular momentum of the meson-like folded string
discussed in section \ref{RotationClosed} As a consequence, the energy spectrum
of the {\sf Y}-shaped baryonic string with the string junction at rest and mesonic
string are related to each other as
\be
E_{{\sf Y}} (\omega) = \frac32 E_{{\sf I}}(\omega)
\, , \qquad
J_{{\sf Y}} (\omega) = \frac32 J_{{\sf I}}(\omega)
\, .
\label{32}
\ee
For $\omega \gg 1$, in the limit of short string \cite{GubKlePol02}, \re{32}
coincides with the known relation between Regge trajectories of mesons and
baryons described by the hadronic QCD string. For $\omega \to 1$, in the limit
of long strings, one calculates the anomalous dimension of the $N=3$ particle
conformal operators as
\be
\gamma_J^{N=3}
=
E_{{\sf Y}}(\omega) - J_{{\sf Y}} (\omega)
=
\frac32 \left[ E_{{\sf I}}(\omega)-J_{{\sf I}}(\omega) \right]
=
\frac32 \gamma_J^{\rm tw-2}
=
3 {\mit\Gamma}_{\rm cusp}(\alpha_s)\ln J
\, .
\label{gamma-AdS}
\ee
We observe that this expression coincides with the upper bound in the spectrum of
the anomalous dimensions of $N = 3$-particle operators, Eq.~\re{bounds}.

We would like to stress that \re{gamma-AdS} corresponds to the {\sf Y}-shaped
string with the string junction at rest. In general, the classical solutions to
the string equations of motion are parameterized by the classical trajectory of
the junction, $X_M = X_M^{\rm (junction)} (\tau)$. For given total angular momentum
$J$ the minimal classically allowed energy of the {\sf Y}-shaped string depends on
the junction trajectories. It is well-known that on the flat background and junction
moving, the minimal energy of the string with the total angular momentum $J =
J_{{\sf Y}}$ is smaller than the energy $E_{{\sf Y}}(\omega)$ defined in \re{32}.
Obviously, the same property holds for short strings on the AdS background. Going
over to the limit of long strings, one expects that the energy levels do not
collide and, as a consequence, the same hierarchy is preserved. In other words,
the minimal energy of the {\sf Y}-shaped long string with the string junction
moving is smaller than the energy with the junction at rest. This implies that
the anomalous dimensions of the corresponding $N = 3$-particle conformal operators
is smaller than the anomalous dimension $\gamma_J^{N = 3}$ defined in \re{gamma-AdS}.
Moreover, the minimal energy of the {\sf Y}-shaped string for the given total
angular momentum $J$ corresponds to the diquark-quark configuration when the
string junction is located near the folding point. In that case, the energy
and the angular momentum of the string approaches the energy and the angular
momentum of the mesonic string, $E_{{\sf I}}(\omega)$ and $J_{{\sf I}}(\omega)$,
respectively, and, as a consequence, the anomalous dimension of the $N = 3$-particle
conformal operator coincides with the twist-two anomalous dimension,
$2 {\mit\Gamma}_{\rm cusp} (\alpha_s) \ln J$. These properties are in a perfect
agreement with the expression obtained before within the Wilson line approach
Eq.~\re{bounds}.

We recall that at $N=3$ the spectrum of the anomalous dimensions, $\gamma_J
(\ell_1)$, is parameterized by integer $0\le \ell_1 \le J$. We have demonstrated
that the two ``extreme'' classical trajectories of the junction, that is the
junction at rest and rotating along the AdS boundary, are mapped into the upper
and lower boundaries of the band, $\ell_1 = 0$ and $\ell_1 = J$, respectively.
We expect that similar correspondence exists for arbitrary $0 < \ell_1 < J$.

\begin{figure}[t]
\unitlength1.5mm
\begin{center}
\mbox{
\begin{picture}(0,35)(20,0)
\put(0,0){\insertfig{6}{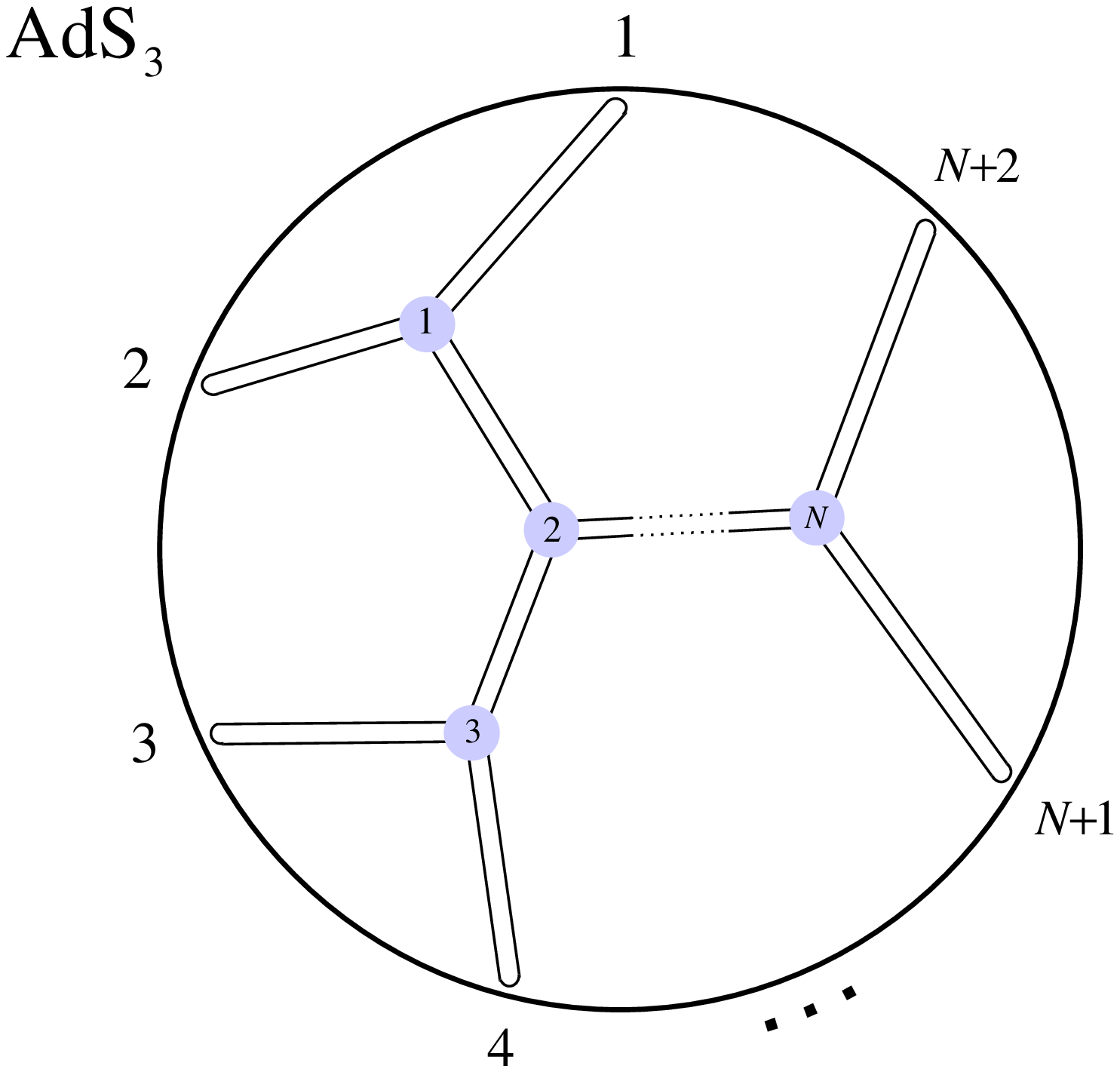}}
\end{picture}
}
\end{center}
\caption{\label{Multiparticle} String picture of multi-particle operators.}
\end{figure}

Let us now consider the $N$-particle conformal operators. As was shown in section
\ref{Section-Multiparticle}, the spectrum of their anomalous dimensions at weak
coupling is parameterized by the set of $N - 2$ integers $\ell_i$. Going over to
the strong coupling limit, we expect that the same structure should be present.
In other words, the string configuration describing such operators has to manifest
the $(N - 2)$ additional degrees of freedom. At $N = 3$ such degree of freedom is
provided by the string junction. For $N \ge 4$ one can use the {\sf Y}-shaped
folded string as a building block to construct the classical string with an
arbitrary number of bits. An example is shown in Fig.\ \ref{Multiparticle}.
Notice that the number of junctions for the string with $N$ folding points
equals $N - 2$.  It worth mentioning that similar configurations in hadronic
QCD string describe exotic mesons and baryons \cite{Art75}.

A general analysis of such string configurations is rather involved, even on the
flat background. Nevertheless, the asymptotic behavior of the anomalous $E - J
\sim N {\mit\Gamma}_{\rm cusp} \ln J$ can be easily derived by considering the
limiting case when all $N$ folding points approach the AdS boundary. As in the
$N = 3$ case, the $(N - 2)$ junctions are at rest so that the energy and angular
momentum of the string with $N$ arms receives an additive contribution from each
arm. This result agrees with the upper bound in the spectrum of the $N$-particle
conformal operator, Eq.~\re{gamma-max}.

A natural question arises about the possible physical interpretation of the string
junction. If the string junction is a genuine physical degree of freedom of
$N$-particle operators, it should be also found at weak coupling. As we have
discussed in section \ref{HeisenbergChain}, the anomalous dimensions at weak
coupling are identified as the eigenvalues of the one-loop QCD dilatation operator
which coincides at large Lorentz spin $J$ with spin chain Hamiltonian,
Eq.~\re{gamma-int}. In the quasiclassical approach, the anomalous dimensions are
calculated by imposing the Bohr-Sommerfeld quantization conditions on the orbits
of classical motion of $N$ particles on the light-cone. The latter can be
significantly simplified by going over from the original, light-cone
$\xi$-coordinates to the separated $z$-coordinates. The Hamilton-Jacobi equations
for the action function $S_0(z)$ take the following form in the separated coordinates
\be
y^2 = t_N (z) - 4 z^{2N}
\label{surface1}
\ee
where $y = 2 z^N \sinh S_0'(z)$ and $t_N(z) = 2 z^N + q_2 z^{N-2} + \ldots + q_N$
with $q_2 = - J^2$ at large $J$ and $q_k$ being the higher integrals of motion.
This hyperelliptic curve of genus $N - 2$ is a ``surface of equal energy'' for
a given set of the integrals of motion $q_k$ which define the coordinates on
the moduli space of the complex structures of the Riemann surface \re{surface1}.
Quasiclassical calculation of the anomalous dimension at weak coupling, Eq.\
\re{gamma-int}, amounts to the quantization of the moduli space of these complex
structures.

In particular, at $N = 3$ the Riemann surface \re{surface1} corresponding to the
baryonic operator has the topology of a torus. In that case, the collective
degrees of freedom ``live'' on this surface. The Bohr-Sommerfeld quantization
conditions allow one to find the quantized values of the integral of motion
$q_3 = q_3 (J,\ell_1)$ and calculate the corresponding anomalous dimension
$\gamma_J (J,\ell_1)$. It turns out that the upper and the lower bounds in their
spectrum, $\gamma_J (J,\ell_1 = 0)$ and $\gamma_J (J,\ell_1 = J)$, correspond to
$q_3^2 = J^6/27$ and $q_3 = 0$, respectively. At these values of $q_3$ the Riemann
surface \re{surface1} becomes degenerate, that is one of the cycles shrinks into a
point. From the point of view of classical mechanics this corresponds to freezing
out the collective degrees of freedom. We observe that the same phenomenon occurs
at the strong coupling. Namely, the string junction is at rest at the upper bound
of the spectrum and at the di-quark center-of-mass at the lower bound of the
spectrum. This suggests that the classical dynamics of the string junction is
governed by yet another Riemann surface of the same genus. Indeed, it is known
that the general solutions to the string equations of motion \re{EQs} are
parametrized by a hyperelliptic curve of higher genus \cite{Kri94}. The explicit
form of this curve is fixed by the boundary conditions. In the case of the string
with $N - 2$ junctions such conditions are given by Eqs.\ \re{EQ-junction}. It
would be interesting to compare \re{surface1} with the curve emerging at strong
coupling.

\section{Concluding remarks}

The present paper was devoted to studies of the anomalous dimensions of
conformal operators at weak and strong coupling. We have demonstrated that
in the both regimes the anomalous dimensions behave asymptotically as
$\sim {\mit\Gamma}_{\rm cusp}(\alpha_s) \ln J$ at large Lorentz spin $J$
while the dependence of the cusp anomalous dimension on the coupling constant
is different. At weak coupling, we calculated the first two terms of
perturbative expansion of ${\mit\Gamma}_{\rm cusp}(\alpha_s)$ in a generic
gauge theory involving scalars. While at large coupling we obtained its
leading asymptotic behavior using classical limit of string propagating on
the AdS background.

In perturbative regime, we have found the one-loop cusp anomaly corresponds to
angular gluon propagation on a cylinder. This allows to establish a relation of
the former to the quantum transition amplitude of a spherical top. Due to
localization phenomena, it is  saturated by classical trajectories, i.e.\
multiple windings of paths around the principle circles of a sphere. All of these
properties are naturally incorporated into the two-dimensional gauge theory on a
disk. There, the cusp anomaly is expressed as an integral of the partition
function with a boundary holonomy with respect to the area of the disk. The
well-known relation of Yang-Mills in two dimensions to a string theory, give us
the opportunity to give a stringy representation of the cusp anomalous dimension
at weak coupling.

At strong coupling, we extended the Gubser-Klebanov-Polyakov results for
twist-two operators as a rapidly rotating closed string to multi-particle
cases. The integrability of one-loop interaction kernel implies that the
$N$-particle anomalous dimension is a function of $N$ parameters, ---
conserved charges. One expects that strong coupling will share similar
property, --- the anomalous dimension will keep this properties. In the
stringy picture, these new degrees of freedom are encoded into the string
junctions. The equation of motion for the latter are parametrized by a
hyperelliptic curve, i.e.\ their classical dynamics of the string is
driven by a Riemann surface. This will be discussed elsewhere.

To establish the relation between the two expressions for ${\mit\Gamma}_{\rm cusp}
(\alpha_s)$ using the gauge/string correspondence, one has to provide the explicit
mapping between the conformal operators in Yang-Mills theory and eigenstates of
the stringy Hamiltonian on some background. Then, one can identify the anomalous
dimensions of the conformal operators for arbitrary $\alpha_s$ as the energy of
the corresponding stringy excitations. To go over from the strong coupling regime
to arbitrary coupling constant in gauge theory, one needs to know the whole
spectrum of the quantum string. At present this problem can not be solved in
full due to lack of the quantization of the strings on AdS${}_5 \times$S${}^5$
background.

It is known that this difficulty can be avoided by considering the Penrose
limit of the AdS${}_5 \times$S${}^5$ background. It is relevant to calculation
of the anomalous dimension of local operators in the ${\cal N} = 4$ YM theory
with large $R$-charge \cite{BerMalNas02}. The string theory on this background
is exactly solvable and, as a consequence, the spectrum of the stringy
excitations can be found. In this case, the gauge/string correspondence looks
as follows. There are six adjoint scalars ${\mit\Phi}_i$ in ${\cal N} = 4$
theory and the $R$-symmetry rotates two of them, say ${\mit\Phi}_1$ and
${\mit\Phi}_2$. The stringy oscillator states are mapped into the so-called
BNM operators constructed from the complex field $Z (x) = {\mit\Phi}_1 (x)
+ i {\mit\Phi}_2 (x)$. Namely, the operator $\tr \, [Z^{J} (0)]$ with the large
$R$-charge $J \gg 1$ is dual to the ground state of the string of the length
$J$, $\ket{0,J}$, while the operator with two ``impurities'' is dual to excited
oscillatory stringy state
\beq
a_n^{i \dagger} a_{-n}^{j \dagger}
\ket{0,J} \leftrightarrow \sum_{l = 0}^J \e^{2 i \pi l n/J}
\tr \, [ {\mit\Phi}_i \, Z^l \, {\mit\Phi}_j \, Z^{J - l} ]
\, ,
\eeq
where $i,j = 1, \ldots, 6$. The exact spectrum of the string Hamiltonian in
the pp-wave background gives rise to the anomalous dimensions of the BMN
operators with large $R$-charge for arbitrary coupling constant. At strong
coupling they coincide with expressions obtained in the quasiclassical
approximation while at weak coupling they match the first few terms of
perturbative expansion \cite{GroMikRoi02,SanZan02}.

It turned out that the quantum string on the pp-background is intrinsically
related to integrable spin chains. The latter appear when one examines the
renormalization of the local operators $\tr \, [ {\mit\Phi}_{i_1} (0) \dots
{\mit\Phi}_{i_k} (0) ]$ at weak coupling in the multi-color limit. These
operators mix under renormalization already at one-loop level and their
eigenvalues can be found by diagonalizing the corresponding mixing matrix
\cite{mixing}. As was found in \cite{MinZar02}, the one-loop mixing matrix
coincides with the Hamiltonian of a completely integrable $SO (6)$ Heisenberg
spin chain defined in an appropriate basis. The length of the spin chain is
equal to $k$ and the spin operators belong to the fundamental representation
of the $SO (6)$ group. The appearance of this group can be traced back to the
fact that the same group is the isometry group of the S${}^5$.

We observe a striking similarity between renormalization properties of such
operators and conformal operators discussed above. In both cases, the one-loop
mixing matrix gives rise to an integrable spin chain. The dynamical symmetry
group of the spin chain, --- the $SO(6)$ group for the local scalar operators
and the $SL(2,\mathbb{R})$ group for the conformal operators, --- is dictated
by the isometry of the relevant part of the background, the $S^5$ and the AdS
parts, respectively. In spite of the fact that two spin chains are different
their energy spectrum can be obtained within the Bethe Ansatz in a similar
manner by quantizing their spectral curves. For the $SL (2,\mathbb{R})$ magnet
the spectral curve, Eq.~\re{surface1}, is hyperelliptic and its genus equals
the number of fields involved. For the $SO(6)$ magnet the curve is more complicated
and it can be reduced to hyperelliptic curve of the genus $J$ if one considers
only its $SO(3)$ subgroup. Having these properties in mind, one may consider a
more general case of renormalization of a local composite operator built from
an arbitrary number of scalar fields and covariant derivatives acting along
different light-cone directions, $(n_j \cdot {\cal D}) {\mit\Phi}_i (0)$ with
$n_j^2 = 0$ and $i = 1, \ldots, 6$. One might expect that the corresponding
one-loop mixing matrix is related to the spin chain with the symmetry group
$SO(2,4) \times SO(6)$, which is the isometry group of the AdS${}_5 \times$S${}^5$
background.

Going over to the strong coupling regime, one should ask about the fate of
integrability of the mixing matrix. For the scalar, BMN like operators it has
been suggested that integrability holds to higher loop orders \cite{BeiKriSta03}.
Would it be the case, the transition of the anomalous dimensions from weak to
strong coupling regime would correspond to the flow in the space of integrable
Hamiltonians with respect to the coupling constant $\alpha_s$. Moreover, had
the same property be valid for conformal operators, it would allow one to
calculate their anomalous dimensions at large spin $J$ for arbitrary $\alpha_s$.
This question certainly deserves further studies.

We would like to stress that the origin of integrability of the one-loop mixing
matrix remains obscure. A possible explanation could come from the stringy
picture for the cusp anomalous dimension as weak coupling discussed in section
\ref{WeakCusp}. We have argued that the corresponding string picture emerges
from the two-dimensional Yang-Mills theory which in its turn is equivalent to
topological theory at $g^2 = 0$ with the gauge group $SL (2,\mathbb{R})$. The
latter theory is the limit of the $SL (2,\mathbb{R})$ Chern-Simons theory at
the level $k \to \infty$. It is known that the correlation functions of the
Wilson lines in the Chern-Simons theory exhibit integrable structure related
to the XXZ Heisenberg spin chain with symmetry group $SL (2,\mathbb{R})$ and
anisotropy $q = \e^{2 \pi i/(k - 2)}$. In this way, for $k \to \infty$ one
recovers the homogeneous XXX spin chain. The use of Chern-Simons approach
for the calculation of the anomalous dimensions of the operators with arbitrary
conformal spins and the corresponding stringy picture behind will be discussed
elsewhere.

\vspace{1cm}

We would like to thank I.~Kogan, Yu.~Makeenko, K.~Zarembo for useful discussions
and E.~Floratos, A.~Kehagias, A.~Kotikov, V.~Velizhanin for correspondence on
their two-loop computation of anomalous dimensions. A.G. thanks TPI at University
of Minnesota, TPI at Uppsala University where a part of the work was done and
IHES where it has been completed for the kind hospitality. This work was supported
by the US Department of Energy under contract DE-FG02-93ER40762 (A.B.) and in part
by grants INTAS-00-334 and RFBR-01-01-00549 (A.G.).

\appendix
\setcounter{section}{0} \setcounter{equation}{0}
\renewcommand{\theequation}{\Alph{section}.\arabic{equation}}

\section{Cusp anomaly in dimensional reduction}
\label{DRED}

The difference between two-loop expressions for the cusp anomalous dimension
in the dimensional regularization and dimensional reduction schemes,
Eqs.~\re{two-loop} and \re{two-loop-DR}, respectively, is solely due to
difference of the corresponding one-loop gluon polarization operators. In
gauge theory it receives contribution from gauge bosons, $n_f$ fundamental
fermions and $n_s$ scalars. In the momentum representation in two different
schemes one obtains (with the Feynman gauge)

$\bullet$ dimensional regularization (DREG):
\begin{equation}
\label{PolarizationDREG}
{\mit\Pi}^{\mbox{\tiny DREG}}_{\mu\nu} (q) =
\frac{\alpha_s}{2 \pi}
\left( \frac{4 \pi \mu^2}{- q^2} \right)^\varepsilon
\frac{
{\mit\Gamma} (\varepsilon) {\mit\Gamma} (1 - \varepsilon) {\mit\Gamma} (2 -
\varepsilon) }{ {\mit\Gamma} (4 - 2 \varepsilon) }
\left( q^2 g^{(d)}_{\mu\nu} - q_\mu q_\nu \right)
\Big[
N_c (5 - 3 \varepsilon) - 2 n_f (1 - \varepsilon) - n_s/2
\Big]
,
\end{equation}

$\bullet$ dimensional reduction (DRED):
\begin{eqnarray}
\label{PolarizationDRED}
&&\!\!\!\!\!\!\!\! {\mit\Pi}^{\mbox{\tiny DRED}}_{\mu\nu} (q) =
\frac{\alpha_s}{2 \pi}
\left( \frac{4 \pi \mu^2}{- q^2} \right)^\varepsilon
\frac{
{\mit\Gamma} (\varepsilon) {\mit\Gamma} (1 - \varepsilon) {\mit\Gamma} (2 -
\varepsilon) }{ {\mit\Gamma} (4 - 2 \varepsilon) }
\\
&&\qquad\times
\bigg\{
q^2 \left( g^{(4)}_{\mu\nu} - g^{(d)}_{\mu\nu} \right) \!
\Big[ N_c - n_f + n_s/2 \Big]
+
\left( q^2 g^{(4)}_{\mu\nu} - q_\mu q_\nu \right) \!
\Big[
N_c (5 - 4 \varepsilon) - 2 n_f (1 - \varepsilon) - n_s/2
\Big]
\bigg\}
, \nonumber
\end{eqnarray}
where $g^{(4)}_{\mu\nu}$ and $g^{(d)}_{\mu\nu}$ are metric tensors in the
Minkowski space-time of dimension $4$ and $d = 4 - 2 \varepsilon$ with
$\varepsilon > 0$, respectively, and $\alpha_s$ is a bare coupling constant.
Note that in supersymmetric theories the $d$-dimensional Lorentz non-covariant
part vanishes in the right-hand side of \re{PolarizationDRED}
\begin{equation}
N_c - n_f + n_s/2 = 0 \, ,
\end{equation}
which is easy to very at ${\cal N} = 1$ ($n_f = N_c$, $n_s = 0$), ${\cal N} = 2$
($n_f = 2 N_c$, $n_s = 2 N_c$) and ${\cal N} = 4$ ($n_f = 4 N_c$, $n_s = 6 N_c$).

The regularized polarization operators in the DREG and DRED schemes have the
same residue at the pole in $\varepsilon$ and differ by a finite $\mathcal{O}
(\varepsilon^0)$-term due to contribution of $\varepsilon$-scalars in the DRED
scheme. To renormalize the obtain expressions, we apply the modified minimal
subtraction procedure. In this way, subtracting the ultraviolet pole from
${\mit\Pi}^{\mbox{\tiny DRED}}_{\mu\nu}$, one defines the so-called ${\rm
\widebar{DR}\!}$ renomalization scheme for the coupling constant. The
counter-term in the ${\rm\widebar{DR}\!}$ scheme is given by
\begin{equation}
{\rm div}_{{\rm \widebar{\scriptscriptstyle DR\!}}} \,
{\mit\Pi}^{\mbox{\tiny DRED}}_{\mu\nu}
=
\frac{
\alpha_s^{{\rm \widebar{\scriptscriptstyle DR\!}}}
}{ 12 \pi }
\left( q^2 g^{(4)}_{\mu\nu} - q_\mu q_\nu \right)
\left( 5 N_c - 2 n_f - n_s/2 \right)
\left( \frac{1}{\varepsilon} - \gamma_{\rm E} + \ln 4 \pi \right)
\,.
\label{ct}
\end{equation}
One can define yet another renormalization scheme by adding a finite term to
the right-hand side of \re{ct}
\begin{equation}
\label{MS-scheme}
{\rm div}_{{\rm \widebar{\scriptscriptstyle MS\!}}} \,
{\mit\Pi}^{\mbox{\tiny DRED}}_{\mu\nu}
=
\frac{
\alpha_s^{_{\rm \widebar{\scriptscriptstyle MS\!}}}
}{ 12 \pi }
\left( q^2 g^{(4)}_{\mu\nu} - q_\mu q_\nu \right)
\left\{
\left( 5 N_c - 2 n_f - n_s/2 \right)
\left( \frac{1}{\varepsilon} - \gamma_{\rm E} + \ln 4 \pi \right)
- N_c
\right\}
\, .
\end{equation}
In this scheme, the renormalized polarization operator ${\mit\Pi}^{\mbox{\tiny
DRED}}_{R} = {\mit\Pi}^{\mbox{\tiny DRED}}-{\rm div}{\mit\Pi}^{\mbox{\tiny DRED}}$
coincides with the polarization operator \re{PolarizationDREG} renormalized
within the conventional dimensional regularization ${\widebar {\rm MS}}$ scheme,
that is ${\mit\Pi}^{\mbox{\tiny DRED}}_{R} = {\mit\Pi}^{\mbox{\tiny DREG}}_{
{\rm \widebar{\scriptscriptstyle MS\!}}}$. That is the reason why one usually
refers to \re{MS-scheme} as the dimensional reduction ${\widebar {\rm MS}}$
scheme. The coupling constants in two schemes are related to each other through
the scheme transformation
\begin{equation}
\alpha_s^{{\rm \widebar{\scriptscriptstyle MS\!}}}
=
\alpha_s^{{\rm \widebar{\scriptscriptstyle DR\!}}}
\lr{
1 -
\frac{N_c}{12}\frac{\alpha_s^{{\rm \widebar{\scriptscriptstyle DR\!}}}}{\pi}
+
\mathcal{O}
\left(
(\alpha_s^{{\rm \widebar{\scriptscriptstyle DR\!}}})^2
\right)
}
\, .
\end{equation}
The polarization operator modifies the gluon propagator by the term
\be
D_{\mu\nu} (x)
\to
D_{\mu\nu}^{(1)}(x)
=
- i
\int\frac{d^d q}{(2 \pi)^d} \e^{- i q \cdot x}
\frac{\mit\Pi_{\mu\nu,\,R} (q)}{q^4}
\, ,
\ee
with ${\mit\Pi}_{R} \equiv {\mit\Pi} - {\rm div} \, {\mit\Pi}$. Its substitution
into \re{W-1-loop} yields the following contribution to the Wilson loop evaluated
along the contour shown in Fig.\ \ref{Cylinder}
\begin{equation}
W^{(1)} = - i g^2 \, t^a t^a
\int \frac{d^d q}{(2 \pi)^d}
\frac{v_\mu {\mit\Pi}_R^{\mu\nu} (q) v'_\nu}{q^4 (q \cdot v) (q \cdot v')}
\, .
\end{equation}
Since the velocity vectors are two-dimensional, $\left( g^{(4)}_{\mu\nu} -
g^{(d)}_{\mu\nu}\right) v_\nu = 0$, so that the first term in Eq.\
(\ref{PolarizationDRED}) does not contribute. Calculating this integral in the
$\widebar{\rm DR\!}$-scheme, one can determine the contribution to the two-loop
cusp anomalous dimension \re{two-loop-DR}, coming from $n_f$ fermions, $n_s$
scalars and part of the $N_c$ term. The remaining terms $\sim N_c$ originate
from other two-loop Feynman diagrams. In the dimensional reduction, the only
difference between $W^{(1)}$ evaluated in the $\rm\widebar{MS}$- and
$\rm\widebar{DR}$-schemes comes from $(-N_c)$ term in the right-hand side of
\re{MS-scheme}. A simple evaluation using the integral
\begin{equation}
\int \frac{d^d q}{(4 \pi)^d}
\frac{v \cdot v'}{[- q^2 + \lambda^2]^m \, (q \cdot v) (q \cdot v')}
=
\frac{2 i}{(4 \pi)^{d/2}} \frac{{\mit\Gamma} (m - d/2 + 1)}{{\mit\Gamma} (m)}
\frac{\theta\coth \theta}{\lambda^{2 m - d + 2}}
\, ,
\end{equation}
with $\lambda^2$ being an infrared cut-off, gives
\begin{equation}
W^{\mbox{\tiny DRED}}_{{\rm \widebar{\scriptscriptstyle DR\!}}}
-
W^{\mbox{\tiny DRED}}_{{\rm \widebar{\scriptscriptstyle MS\!}}}
=
\left(
\frac{
\alpha_s^{{\rm \widebar{\scriptscriptstyle DR\!}}}
}{
\pi
}
\right)^2
\frac{C_F N_c}{12} \, \theta \coth \theta \, \ln \frac{\mu}{\lambda}
\, .
\label{diff-W}
\end{equation}
Notice that, by construction, $W^{\mbox{\tiny DRED}}_{{\rm
\widebar{\scriptscriptstyle MS\!}}} = W^{\mbox{\tiny DREG}}_{{\rm
\widebar{\scriptscriptstyle MS\!}}}$ to two-loop level. At large $\theta$,
Eq.~\re{diff-W} is translated into similar relation between the cusp
anomalous dimension in two schemes, Eqs.~\re{two-loop} and \re{two-loop-DR}.


\end{document}